\documentclass[aip,cha, reprint,showpacs,showkeys]{revtex4-1}
\usepackage{amssymb, dcolumn,  graphicx, latexsym, amsfonts, bm}

\begin{document}

\title{Ion cyclotron parametric turbulence and anomalous convective transport of the inhomogeneous plasma in front of the fast wave antenna}

\author{V. S. Mikhailenko}\email[E-mail:]{vsmikhailenko@pusan.ac.kr}
\affiliation{Plasma Research Center, Pusan National University,  Busan 46241, South Korea}
\author{V. V. Mikhailenko}\email[E-mail: ]{vladimir@pusan.ac.kr}
\affiliation{BK21 FOUR Information Technology, Pusan National University,  Busan 46241, South Korea.}
\author{Hae June Lee}\email[E-mail: ]{haejune@pusan.ac.kr}
\affiliation{Department of Electrical Engineering, Pusan National University, Busan 46241, South Korea}
\date{\today}

\begin{abstract}
The theory of the ion cyclotron (IC) electrostatic parametric instabilities of the inhomogeneous plasma which 
are driven by the inhomogeneous electric field of the fast wave (FW) in front of FW antenna is developed. It is found that the spatial inhomogeneity of the 
parametric IC turbulence and of the drift turbulence is at the origin of the flows which transport a large part of FW power, 
deposited to the pedestal plasma, to the scrape-off layer and later to the divertor mostly along field lines.

\end{abstract}

\maketitle

\section{Introduction}\label{sec1}

Fast wave (FW) heating in the ion cyclotron range of frequency (ICRF)\cite{Stix} is used in the current-state magnetic fusion devices 
and will be applied in the future devices, like ITER, as one of the main methods of the auxiliary heating and sustaining H-mode 
tokamak plasmas. The non-inductive current drive with FWs at frequencies far above the ion cyclotron frequency and approaching the lower hybrid frequency 
(also called 'helicons' or 'whistlers') is considered\cite{Hosea,Prater,Wang,Pinsker} to be essential to sustain steady state plasma operation 
in tokamaks to complement the bootstrap 
current. Last decades, essentially all experiments with FW heating of the tokamak plasmas have found that up to 50$\%$ or more of the
launched RF power can be lost in the region outside of the last closed flux surface (LCFS) in the vicinity of the antenna in the scrape-off 
layer (SOL) - the region of the plasma between LCFS and tokamak vessel. It is obvious that so large RF power losses in the SOL can have 
a significant impact on the performance of the FW heating in the current magnetic fusion experiments and for the long pulse plasmas 
anticipated in the ITER experiment. 

The usually applied linear theory of the propagation and absorption of FW in a weakly inhomogeneous plasma gives proper description of the FW plasma heating 
in the core plasma bounded by LCFS, where the quiver velocity of ions and electrons in FW is negligible in comparison 
with the thermal velocities of the plasma components. This linear theory, however, is not applicable to the treatment of the interaction of FW with a 
plasma near FW antenna, where the ion quiver velocity is not negligibly small and may be commensurable with ion thermal velocity. Therefore, 
in the near - antenna region linear physics fails and the anomalously strong absorption of the RF energy in the near - antenna region is attributed to the 
development of the nonlinear processes. 

It was generally observed that a significant heating of the low energy ions in the SOL 
have been regularly accompanied by the development of different parametric instabilities. It was claimed in Refs. \cite{Rost,Pace, Wilson} 
that the observed anomalous heating of the SOL ions and absorption by this means of the FW energy is the result of interactions of ions
with the ion cyclotron (IC) parametric quasi-mode decay instabilities\cite{Porkolab1, Porkolab2} which were detected in the near antenna region. 

Large efforts were applied to the development of the theory of the parametric instabilities driven by the strong FW\cite{Porkolab1, 
Porkolab2,Mikhailenko1,Mikhailenko2,Mikhailenko3}. These instabilities 
stem from the relative oscillatory motion of the plasma species in a plasma under the action of FW. 
It is usually assumed in this theory that the 
approximation of the spatially homogeneous FW (dipole approximation) oscillating with time as $\sim \cos \omega_{0}t$, or 
of a plane wave structure $\sim \exp\left(i\mathbf{k_{0}r}-i\omega_{0} t\right)$ of FW are sufficient for the treatment of the parametric instabilities,
since the wave number of the excited waves is much larger than that of the pump wave. This theory is valid for the spatially uniform plasma, where the 
spatial variation of the FW wavenumber $\mathbf{k_{0}}$ is insignificant and the displacement of particles in such a wave is much less than the FW 
wavelength.  The application of such a linear and nonlinear theory 
to the  SOL plasma reveals \cite{Mikhailenko3} that the experimentally observed anisotropy of the heating of cold SOL ions\cite{Wilson} may be 
attributed to the interaction of the SOL ions with IC parametric turbulence. However, this heating is very weak and is not commensurable with much more 
stronger anomalous ion heating observed in the near-SOL region close to the separatrix where plasma temperature and density gradients are the largest. 
Therefore the clear understanding the role of the observed parametric instabilities in the FW power losses in SOL is still missing and 
there is still lack a verification by the theory.

The FW heating experiments on the National Spherical Torus eXperiment (NSTX) showed that significant losses of the FW energy occur directly in the SOL.
With a fast visible color camera view of the NSTX plasma it was discovered\cite{Perkins, Perkins1} that a large part of RF power loss flows through the SOL 
to the divertor mostly along field lines and is deposited in bright spirals on the divertor. It was found that these flows originate not on the 
antenna but throughout the SOL in front of the antenna. Infrared camera measurements of these spirals show a significant RF-produced heat 
flux transported in these spirals. It was concluded\cite{Perkins, Perkins1} that the underlying mechanism, still unknown, for this loss of 
power is FW propagation in the SOL and that this effect is likely to be present to some degree for FW heating generally. 

Recently, the full wave code AORSA (all-orders spectral algorithm)\cite{Jaeger, Green}, in which the edge plasma beyond LCFS is included
in the solution domain, was employed \cite{Bertelli,Bertelli1} for the simulations of the FW propagation and the FW power loss in SOL.
In this code, a collisional damping parameter is used as a proxy to represent 
the real, and most likely nonlinear, processes of the FW damping in SOL. 3D AORSA results for the NSTX have showed\cite{Bertelli,Bertelli1} 
a direct correlation between the location of the fast wave cut-off layer, the large amplitude of the RF fields in the scrape-off region, 
and the power transition to higher SOL power losses (driven by the RF field). A strong growth of the RF field near antenna was observed 
when the FW cut-off is removed in front of the antenna by 
increasing the edge density. High density in front of the antenna, although positive for the antenna-plasma coupling, leads to an increase of 
the RF electric field in the SOL and corresponding RF power losses\cite{Bertelli}.

These experimental results and numerical simulations reveal that the observed strong FW power loss occurs 
in the region where the plasma density and the FW field amplitude strongly change on the distance much less than the spatial 
inhomogeneity length of the FW field in the bulk of the tokamak plasma. This 
spatial scale, however, is much larger than the wavelengths of the IC parametric turbulence perturbations. The existing theory  of 
the parametric instabilities mentioned above and the theory of the anomalous absorption of the FW energy do not distinguish this 
intermediate scale (mesoscale) and therefore do not reveal the processes which occurs on the spatial scale commensurable with lengths 
of the strong inhomogeneities of the plasma density and the ion temperature in the pedestal and the adjacent near-SOL regions and of 
the local FW electric field inhomogeneity in these regions. 

The focus of this paper is the theory of the IC parametric instabilities and of the anomalous plasma transport in the inhomogeneous FW electric field in 
front of FW antenna. We use a slab geometry with the mapping $\left(r,\theta, \zeta\right)\rightarrow \left(\hat{x}, \hat{y}, \hat{z}\right)$, where $r,
\theta, \zeta$ are the radial, poloidal and toroidal directions, respectively, of the toroidal coordinate system. 
Our investigations of the IC parametric turbulence in the present paper bases on the methodology of the oscillating modes, developed in Ref.
\cite{Mikhailenko1} for the spatially uniform RF electric field, and extended in Ref.\cite{Mikhailenko3} on the FW electric field 
of a plane wave structure with a finite wave length. According to the experimental\cite{Perkins, Perkins1} and numerical \cite{Bertelli,Bertelli1} 
results the plasma density, the ion temperature and FW electric field $\mathbf{E}_{0}\left(\mathbf{\hat{r}}, \hat{t}\right)$ are inhomogeneous 
along radial coordinate $\hat{x}$ directed from FW antenna to bulk of plasma.
Because FW electric field inhomogeneity scale length $L_{E}$ along coordinate $\hat{x}$ in front of antenna is less than FW wavelengths 
$2\pi/k_{0y}$ and $2\pi/k_{0z}$, we can't use in our theory the plane wave model for FW considered in Ref.\cite{Mikhailenko3}. Instead, 
we employ for FW electric field in this region the simple approximation of the oscillating, but spatially inhomogeneous electric field, 
\begin{eqnarray}
&\displaystyle \mathbf{E}_{0}\left(\mathbf{\hat{r}}, \hat{t}\right)=\mathbf{E}_{0x}\left(\hat{x}\right)\cos \omega_{0}\hat{t}+ 
\mathbf{E}_{0y}\left(\hat{x}\right)\sin \omega_{0}\hat{t}.
\label{1}
\end{eqnarray}
It is assumed that cut-off is absent from in front of antenna region, which involves the pedestal and the adjacent near-SOL, due to the 
high increased plasma density in this region\cite{Bertelli,Bertelli1,Green}. 
For the numerical data presented in Refs.\cite{Bertelli,Bertelli1} the oscillatory velocity of 
ions in this region is commensurable with their thermal velocity. In this case the nonlinear FW-plasma interaction, which 
was not investigated yet, can alter the structure and the spatial profiles $\mathbf{E}_{0x}\left(\hat{x}\right)$ and $\mathbf{E}_{0y}
\left(\hat{x}\right)$ of FW electric field, predicted by the linear theory\cite{Vdovin}, and remain only partially understood for this region. 

In this paper, we develop the theory of the microscale IC parametric instabilities driven by the inhomogeneous FW field (1) and 
their mesoscale effects on the inhomogeneous plasma. In Sec. \ref{sec2}, we present for the first time the basic equations and transformations of the 
oscillating modes approach applicable to the inhomogeneous FW electric field (\ref{1}). In view of very different 
time and spatial scales involved in this problem we split our task into two problems. The first problem, considered in Sec. \ref{sec3}, 
describes the fast time scale dynamics of the short wavelength IC instabilities in the inhomogeneous FW. The processes of the IC instabilities 
excitation and saturation and the process of the turbulent heating of plasma components, considered in this sections, occur locally on small spatial 
distances which are much less than the spatial inhomogeneity lengths of plasma parameters 
and of FW field. This theory bases on the local approximation, analytically justified in Ref.\cite{Mikhailenko3}
for the plane wave model for FW, and, developed in Sec. \ref{sec2} for the inhomogeneous FW field (\ref{1}).

In Sec. \ref{sec4}, we make a step beyond  this local theory of the IC parametric turbulence. We develop for the first time the kinetic theory of the 
mesoscale spatial and temporal evolution of the plasma caused by the spatially inhomogeneous short wavelength IC turbulence. We found that 
the spatial inhomogeneity of the IC parametric microturbulence, originated from the spatial inhomogeneity of the plasma and FW field, is at the origin of 
the mesoscale flows, which have spatial scale commensurable with scales of the plasma and FW inhomogeneities. Conclusions are given in Sec. \ref{sec5}.

\section{Basic equations and transformations}\label{sec2}
Our theory bases on the Vlasov equation for the velocity distribution function $F_{\alpha}$ of $\alpha$ plasma 
species ($\alpha = i$ for ions and $\alpha = e$ for electrons),
\begin{eqnarray}
&\displaystyle \frac{\partial F_{\alpha}}{\partial
t}+\left. \mathbf{\hat{v}}\frac{\partial F_{\alpha}}
{\partial\mathbf{\hat{r}}}+\frac{e}{m_{\alpha}}\right(\mathbf{E}_{0}\left(\hat{x}, t\right)
\nonumber 
\\ 
&\displaystyle
\left.+\frac{1}{c}\left[\mathbf{\hat{v}}\times
\mathbf{B}\right]-\bigtriangledown
\varphi\left(\mathbf{\hat{r}},t\right)\right)\frac{\partial
F_{\alpha}}{\partial\mathbf{\hat{v}}}=0,\label{2}
\end{eqnarray}
and on the Poisson equation for the self-consistent electrostatic potential $\varphi
\left(\mathbf{\hat{r}}, t\right)$ of the plasma respond on FW,
\begin{eqnarray}
&\displaystyle 
\vartriangle \varphi\left(\mathbf{\hat{r}},t\right)=
-4\pi\sum_{\alpha=i,e} e_{\alpha}\int f_{\alpha}\left(\mathbf{\hat{v}},
\mathbf{\hat{r}}, t \right)d{\bf \hat{v}}. \label{3}
\end{eqnarray}
The magnetic field $\mathbf{B}=\mathbf{B}_{0}+\mathbf{B}_{1}\left(\hat{x}, t\right)$, where $\mathbf{B}_{0}$ and $\mathbf{B}_{1}\left(\hat{x}, t\right)$ 
are the uniform plasma-confining and oscillating FW magnetic fields, respectively, both directed along coordinate $\hat{z}$. For the electric field (\ref{1}), 
\begin{eqnarray}
&\displaystyle 
\mathbf{B}_{1}\left(\hat{x}, t\right)=\frac{c}{\omega_{0}}\frac{dE_{0y}\left(\hat{x}\right)}{d \hat{x}}\cos \omega_{0}t\,\mathbf{e}_{z}.
\label{4}
\end{eqnarray}
In Eq. (\ref{2}), $f_{\alpha}$ is the fluctuating part of the distribution function 
$F_{\alpha}$, $f_{\alpha}=F_{\alpha}-F_{0\alpha}$, where $F_{0\alpha}$ is the equilibrium distribution function. 

Equation (\ref{2}) contains two disparate spatial scales.  It is the mesoscale  which involves in the distribution function $F_{0\alpha}$ and determine the 
spatial inhomogeneity of plasma parameters, and in FW fields, 
determining the spatial inhomogeneity of $\mathbf{E}_{0}\left(\hat{x}, t\right)$ and $\mathbf{B}_{1}\left(\hat{x}, t\right)$. 
The microscale determines the potential $\varphi\left(\mathbf{\hat{r}},t\right)$ and $f_{\alpha}$. It was found in Ref.\cite{Mikhailenko3}, that by the 
transformation of the Vlasov equation to the convective frame we can derive the Vlasov equation for plasma species, in which the macroscales 
may be exclude, as it was for the dipole approximation, or are of on order of small ratio of the small to large spatial scales as for the plane wave 
structure of FW considered in Ref. \cite{Mikhailenko3}.
 
In this paper, we develop this approach for the case of the 
spatially inhomogeneous electric field (\ref{1}). For this goal, we transform the position vector $\hat{\mathbf{r}}$ and velocity $\hat{\mathbf{v}}$ in 
Eq. (\ref{2}) for $F_{\alpha}$, determined in the laboratory frame, to the position vector $\mathbf{r}_{\alpha}$ and velocity $\mathbf{v}_{\alpha}$ 
determined in the frame moving with velocity $\mathbf{V}_{\alpha}\left(\mathbf{r}_{\alpha}, t\right)$.
For the ion Vlasov equation (\ref{2}) the new position vector $\mathbf{r}_{i}=\left(x_{i}, y_{i}, z_{i}\right)$ 
and velocity $\mathbf{v}_{i}=\left(v_{ix}, v_{iy}, v_{iz}\right)$, are determined by the relations
\begin{eqnarray} 
&\displaystyle
\hat{t}=t,\qquad \hat{\mathbf{v}}=\mathbf{v}_{i}+\mathbf{V}_{i}\left(\mathbf{r}_{i}, t\right), 
\nonumber 
\\ &\displaystyle
\hat{\mathbf{r}}=\mathbf{r}_{i}+\mathbf{R}_{i}\left(\mathbf{r}_{i}, t\right)=\mathbf{r}_{i}+\int\limits^{t}
_{t_{(0)}}\mathbf{V}_{i}
\left(\mathbf{r}_{i}, t_{1}\right)dt_{1},\label{5}
\end{eqnarray}
or
\begin{eqnarray}
&\displaystyle
\mathbf{v}_{i}=\hat{\mathbf{v}}-\hat{\mathbf{V}}_{i}\left(\hat{\mathbf{r}},
\hat{t}\right), \,\,\,\mathbf{r}_{i}=\hat{\mathbf{r}}-\int\limits^{\hat{t}}_{t_{(0)}}\hat{\mathbf{V}}_{i}
\left(\hat{\mathbf{r}}, \hat{t}_{1}\right)d\hat{t}_{1},\label{6}
\end{eqnarray}
In new variables, the Vlasov equation for $F_{i}\left(t,\mathbf{r}_{i},\mathbf{v}_{i}\right)$ has a form
\begin{eqnarray}
&\displaystyle 
\frac{\partial F_{i}\left(\mathbf{v}_{i}, \mathbf{r}_{i}, t\right)} {\partial
t}+v_{ij}\frac{\partial F_{i}\left(\mathbf{v}_{i}, \mathbf{r}_{i}, t\right)} {\partial
r_{ij}}-\left(v_{ij}+ V_{ij}\left(
\mathbf{r}_{i}, t\right)\right)
\nonumber 
\\ &\displaystyle
\times\int\limits^{t}_{t_{(0)}}\frac{\partial \hat{V}_{ik}\left(
\hat{\mathbf{r}}, \hat{t}_{1}\right)} {\partial \hat{r}_{j}}d\hat{t}_{1}\frac{\partial
F_{i}\left(\mathbf{v}_{i}, \mathbf{r}_{i}, t\right)}
{\partial r_{ik}}
\nonumber 
\\ &\displaystyle
-v_{ij}\frac{\partial \hat{V}_{ik} \left(\hat{\mathbf{r}},
\hat{t}_{1}\right)}{\partial \hat{r}_{j}}\frac{\partial F_{i}\left(\mathbf{v}_{i}, \mathbf{r}_{i}, t\right)}{\partial
v_{ik}}+\frac{e_{i}}{m_{i}c}\left[\mathbf{v}_{i}
\times\mathbf{B}\right]\frac{\partial
F_{i}}{\partial\mathbf{v}_{i}}
\nonumber 
\\ &\displaystyle
-\left\{ \left[\frac{\partial \hat{\mathbf{V}}_{i}}{\partial\hat{t}}+\hat{\mathbf{V}}_{i}\frac{\partial \hat{\mathbf{V}}_{i}}{\partial \hat{\mathbf{r}}}
\right.\right.
\nonumber 
\\ &\displaystyle
\left.\left.-\frac{e_{i}}{m_{i}}\left(\mathbf{E}_{0}\left(\hat{x}, t\right)
+\frac{1}{c}\left[\hat{\mathbf{V}}_{i}\times\mathbf{B}\right]
\right) \right]\right. 
\nonumber 
\\ &\displaystyle
\left.+\frac{e_{i}}{m_{i}}
\nabla\varphi\left(\hat{\mathbf{r}},\hat{t}\right)\right\}\frac{\partial F_{i}\left(\mathbf{v}_{i}, \mathbf{r}_{i}, t\right)}{\partial \mathbf{v}_{i}} =0.\label{7}
\end{eqnarray}
We define the velocity $\mathbf{\hat{V}}_{i}\left(\mathbf{r},t\right)$  as a function for which the expression in square brackets of 
Eq. (\ref{7}) vanishes, 
\begin{eqnarray}
&\displaystyle 
\frac{\partial \hat{\mathbf{V}}_{i}}{\partial\hat{t}}+\hat{\mathbf{V}}_{i}\frac{\partial \hat{\mathbf{V}}_{i}}{\partial \hat{\mathbf{r}}}
=\frac{e_{i}}{m_{i}}\left(\mathbf{E}_{0}\left(\hat{x}, t\right)
+\frac{1}{c}\left[\hat{\mathbf{V}}_{i}\times\mathbf{B}\right]
\right).\label{8}
\end{eqnarray}
This equation is the Euler equation for the velocity $\hat{\mathbf{V}}_{i}$ of the ideal fluid in the electric $\mathbf{E}_{0}\left(\hat{x}, t\right)$ 
and magnetic $\mathbf{B}$ fields, in which  $\hat{\mathbf{r}}$ is the independent variable. The nonlinear convective derivative in Eq. (\ref{8}) is excluded 
in coordinates  $\mathbf{r}_{i}, t$, determined by Eq. (\ref{8}), with which Eq. (\ref{8}) transforms to the ordinary differential equation for velocity 
$\mathbf{V}_{i}\left(\mathbf{r}_{i},  t\right)$, 
\begin{eqnarray}
&\displaystyle 
\frac{d \mathbf{V}_{i}}{dt}
=\frac{e_{i}}{m_{\alpha}}\left(\mathbf{E}_{0}\left(x_{i}+R_{i}\left(x_{i}, t\right), t\right)
+\frac{1}{c}\left[\mathbf{V}_{i}\times\mathbf{B}\right]\right.
\nonumber 
\\
&\displaystyle
\left.+\frac{1}{c}\left[\mathbf{V}_{i}\times\mathbf{B}_{1}\left(x_{i}+R_{i}\left(x_{i}, t\right), t\right)\right]\right),
\label{9}
\end{eqnarray}
where 
\begin{eqnarray}
&\displaystyle 
R_{i}\left(x_{i}, t\right)=\int \limits ^{t}_{0}dt_{1}V_{ix}\left(x_{i}, t_{1}\right).
\label{10}
\end{eqnarray}
In the ordinary differential equation (\ref{9}), the position vector $\mathbf{r}_{i}$ is a parameter. For distinguishing in what follows the meso- and micro- spatial dependences, 
the variable $X_{i}$ will determine mesoscale spatial variations along $x_{i}$, whereas $\mathbf{r}_{i}=\left(x_{i}, y_{i}, z_{i}\right)$ will stand for the 
microscale variable. Therefore, in projections, Eq. (\ref{9}) becomes

\begin{eqnarray}
&\displaystyle 
\frac{dV_{ix}\left(X_{i}, t\right)}{dt}=\omega_{ci}V_{iy}\left(X_{i}, t\right)
\nonumber 
\\
&\displaystyle
+\frac{e_{i}}{m_{i}c}V_{iy}\left(X_{i}, t\right)B_{1}\left(X_{i}+R_{i}\left(X_{i}, t\right), t\right)
\nonumber 
\\
&\displaystyle
+\frac{e_{i}}{m_{i}}E_{0x}\left(X_{i}+R_{i}\left(X_{i}, t\right), t\right),
\label{11}
\end{eqnarray}
\begin{eqnarray}
&\displaystyle 
\frac{dV_{iy}\left(X_{i}, t\right)}{dt}=-\omega_{ci}V_{ix}\left(X_{i}, t\right)
\nonumber 
\\
&\displaystyle
-\frac{e_{i}}{m_{i}c}V_{ix}\left(X_{i}, t\right)B_{1}\left(X_{i}+R_{i}\left(X_{i}, t\right), t\right)
\nonumber 
\\
&\displaystyle
+\frac{e_{i}}{m_{i}}E_{0y}\left(X_{i}+R_{i}\left(X_{i}, t\right), t\right).
\label{12}
\end{eqnarray}

In Appendix A, we derived the solution (\ref{A3}), (\ref{A4}) of Eqs. (\ref{11}), (\ref{12}) for  $V_{ix}\left(X_{i}, t\right)$ and 
$V_{iy}\left(X_{i}, t\right)$ in the limit of small displacement $|R_{i}\left(X_{i}, t\right)|$ of ion in the FW compared with the electric field 
inhomogeneity scale length $L_{E}$. We employ these solutions to Eq. (\ref{7}) for derivation the explicit form  of (\ref{A9}) for 
$F_{i}\left(\mathbf{v}_{i}, \mathbf{r}_{i}, X_{i}, t\right)$, in which the mesoscale variable $X_{i}$ and the microscale variable 
$\mathbf{r}_{i}$, are included. Equation (\ref{A9}) reveals that because  FW electric 
field is included in the Vlasov equation (\ref{7}) for $F_{i}\left(\mathbf{v}_{i}, \mathbf{r}_{i}, X_{i}, t\right)$ only in the terms which 
contains the $V_{ix}\left(X_{i}, t\right)$  and $V_{iy}\left(X_{i}, t\right)$ velocities gradients, these terms are of the order 
of $|R_{i}/L_{E}|\ll 1$ relative to other terms in the Vlasov equation and should be neglected in Eq. (\ref{7}).
Note, that for the numerical data from Ref.\cite{Bertelli} these terms  are on the order of $2.5\cdot 10^{-2}\ll 1$.  
Without these terms, the Vlasov equation (\ref{A9}) has a form as in the plasma without FW field,
\begin{eqnarray}
&\displaystyle 
\frac{\partial F_{i}\left(\mathbf{v}_{i}, \mathbf{r}_{i}, X_{i}, t\right)} {\partial t}+\mathbf{v}_{i}\frac{\partial F_{i}} {\partial
\mathbf{r}_{i}}+\frac{e_{i}}{m_{i}c}\left[\mathbf{v}_{i}\times\mathbf{B}_{0}\right]\frac{\partial F_{i}}{\partial\mathbf{v}_{i}}
\nonumber 
\\ 
&\displaystyle
-\frac{e_{i}}{m_{i}}\frac{\partial\varphi\left(\mathbf{r}_{i},t\right)}{\partial
\mathbf{r}_{i}}\frac{\partial F_{i}\left(\mathbf{v}_{i}, \mathbf{r}_{i}, X_{i}, t\right)}{\partial \mathbf{v}_{i}} =0.
\label{13}
\end{eqnarray}
The same conclusions are valid for the Vlasov equation (\ref{2}) for the electron distribution function 
$F_{e}\left(\mathbf{v}_{e}, \mathbf{r}_{e}, t\right)$, where $\mathbf{r}_{e}$ and $\mathbf{v}_{e}$ are the position vector and the 
velocity of an electron determined in the frame oscillating relative to the laboratory frame with velocities  $U_{ex}\left(x_{e}, t\right)$ and 
$U_{ey}\left(x_{e}, t\right)$ given by Eqs. (\ref{A5}), (\ref{A6}) with species index $i$ changed on $e$. 

The Vlasov equation (\ref{13}) for the ion distribution $F_{i}\left(\mathbf{v}_{i},\mathbf{r}_{i}, X_{i}, t\right)$, determined in the oscillating 
ion convective frame, the Vlasov equation for the electron distribution $F_{e}\left(\mathbf{v}_{e},\mathbf{r}_{e}, X_{e}, t\right)$, determined 
in the oscillating electron convective frame, and the Poisson equation (\ref{3}) for the self-consistent electrostatic potential $\varphi$ compose the 
governing system of equations in our theory of IC parametric instabilities of the inhomogeneous plasma driven by the inhomogeneous FW field
(\ref{1}). It follows from Eq. (\ref{13}) that the equilibrium ion distribution function $F_{i0}\left(\mathbf{v}_{i}, X_{i}\right)$, which is the 
solution to Eq. (\ref{13}) for $\varphi\left(\mathbf{r}_{i},t\right)=0$, does not contain 
the temporal and spatial dependences caused by the FW field and $F_{i0}\left(\mathbf{v}_{i}, X_{i}\right)$ will be assumed to be the Maxwellian 
distribution,
\begin{eqnarray}
&\displaystyle
F_{i0}\left(\mathbf{v}_{i}, X_{i}, t\right)=\frac{n_{0i}}{\left(2\pi v^{2}_{Ti}\right)^{3/2}}
\exp \left(-\frac{v^{2}_{i}}{2v^{2}_{Ti}}\right),
\label{14}
\end{eqnarray}
where $v^{2}_{Ti}=T_{i}/m_{i}$ is the ion thermal velocity. The equilibrium ion density, $n_{0i}=n_{0i}\left(X_{i}\right)$, and the ion temperature, 
$T_{i}=T_{i}\left(X_{i}\right)$, are assumed to be inhomogeneous along the coordinate $X_{i}$. The same Maxvellian distribution with species index 
$i$ replaced by $e$ is assumed for the electron equilibrium distribution $F_{e0}\left(\mathbf{v}_{e}, X_{e}\right)$. We assume, that the gradient 
scale lengths of the plasma density, $L_{n} = \left[ d\ln n_{0}\left(X_{\alpha} \right)/dX_{\alpha} \right]^{-1}$, of the ion 
temperature, $L_{T} = \left[ d\ln T_{i}\left(X_{i}\right)/dX_{i} \right]^{-1}$, and of the FW electric field $L_{E}=\left[ d\ln E_{0x}\left(X_{i}\right)/
dX_{i} \right]^{-1}$ are much larger than the wavelength of the parametric IC instabilities, 
i. e. $k_{ix}L_{n} \gg 1$, $k_{ix}L_{T} \gg 1$, and $k_{ix}L_{E} \sim L_{E}/|R_{i}| \gg 1$, where $k_{ix}$ is the wave number conjugate with coordinate 
$x_{i}$. These conditions determine the validity of the local approximation in our theory. In Section \ref{sec3}, we develop the theory of the 
microscale IC parametric instabilities of the inhomogeneous plasma in the inhomogeneous FW. In Section \ref{sec4}, we develop the theory of 
the mesoscale spatial evolution of a plasma under the action of the spatially inhomogeneous IC parametric microturbulence. 

\section{The microscale IC parametric instabilities and turbulent heating of the inhomogeneous plasma in the inhomogeneous 
FW electric field}\label{sec3}

The relative oscillatory motion of ions and electrons in the strong RF field is at the origin of the microscale parametric electrostatic instabilities, 

It follows from Eq. (\ref{13}) that the Vlasov equation for the perturbation $f_{i}\left(\mathbf{v}_{i},\mathbf{r}_{i}, X_{i}, t \right)$ 
of the ion distribution function $F_{i0}\left(\mathbf{v}_{i}, X_{i}\right)$ and the similar equation for the perturbation 
$f_{e}\left(\mathbf{v}_{e},\mathbf{r}_{e}, X_{e}, t \right)$ of the 
electron distribution function $F_{e0}\left(\mathbf{v}_{i}, X_{e}\right)$ are the same as for a plasma without FW field,
\begin{eqnarray}
&\displaystyle 
\frac{\partial f_{i}}{\partial t}+\mathbf{v}_{i}\frac{\partial f_{i}}{\partial\mathbf{r}_{i}}
+\frac{e_{i}}{m_{i}c}\left[\mathbf{v}_{i}\times\mathbf{B}_{0}\right]\frac{\partial f_{i}}{\partial\mathbf{v}_{i}}
\nonumber 
\\ 
&\displaystyle
-\frac{e_{i}}{m_{i}}\nabla \varphi_{i}\left(\mathbf{r}_{i},t\right)\frac{\partial
f_{i}}{\partial\mathbf{v}_{i}} =\frac{e_{i}}{m_{i}}\nabla \varphi_{i}\left(\mathbf{r}_{i},t\right)\frac{\partial 
F_{i0}\left(\mathbf{v}_{i}, X_{i}\right)}{\partial\mathbf{v}_{i}},
\label{15}
\\
&\displaystyle 
\frac{\partial f_{e}}{\partial t}+\mathbf{v}_{e}\frac{\partial f_{e}}{\partial\mathbf{r}_{e}}
+\frac{e}{m_{e}c}\left[\mathbf{v}_{e}\times\mathbf{B}_{0}\right]\frac{\partial f_{e}}{\partial\mathbf{v}_{e}}
\nonumber 
\\ 
&\displaystyle
=\frac{e}{m_{e}}\nabla \varphi_{e}\left(\mathbf{r}_{e},t\right)\frac{\partial
F_{e0}\left(\mathbf{v}_{e}, X_{e}\right)}{\partial\mathbf{v}_{e}}.
\label{16}
\end{eqnarray}
In Eqs. (\ref{15}), (\ref{16}), the mesoscale spatial variables $X_{i}$ and $X_{e}$ are involved in $F_{i0}\left(\mathbf{v}_{i}, X_{i}\right)$ and 
$F_{i0}\left(\mathbf{v}_{i}, X_{i}\right)$ and present in these equations as a parameters. The solutions to Eqs. (\ref{15}) and (\ref{16}) 
were derived in Ref.\cite{Mikhailenko3}, where the local approximation to the theory of the parametric IC microinstabilities was developed.
These results for $f_{i}$ and $f_{e}$, as well as ones for the perturbed ion and electron densities, are applicable to the present investigations of the IC 
microinstabilities. We should keep in mind the importance of the mesoscale variables $X_{i}$ and $X_{e}$ which are the parameters in the local 
theory of the microscale instabilities, but become the full-fledged spatial variables in our analyse of the mesoscale evolution 
of the inhomogeneous plasma under the action of the spatially inhomogeneous IC turbulence presented in the next section. 

In this paper, as well as in Ref.\cite{Mikhailenko3}, the Poisson equation (\ref{3}) is investigated as the equation for $\varphi_{i}\left(\mathbf{k}_{i},t
\right)$ by the Fourier transform of Eq. (\ref{3}) over microscale $\mathbf{r}_{i}$, 
\begin{eqnarray}
& \displaystyle 
k^{2}_{i}\varphi_{i}\left(\mathbf{k}_{i},X_{i}, t\right)
=4\pi e\left(\delta n_{i}\left(\mathbf{k}_{i}, X_{i}, t\right)\right.
\nonumber 
\\ 
&\displaystyle
\left.-\int d\mathbf{r}_{i}\delta n_{e}\left(\mathbf{r}_{e}, X_{i}, t\right)e^{-i\mathbf{k}_{i}\mathbf{r}_{i}}\right).
\label{17}
\end{eqnarray}
In Eq. (\ref{17}), the Fourier transform over $\mathbf{r}_{i}$ should be determined for $n_{e}
\left(\mathbf{r}_{e}, X_{i}, t\right)$ as well as for the potential $\varphi_{e}\left(\mathbf{r}_{e}, X_{i}, t_{1} 
\right)$, which is included in the expression for $n_{e}\left(\mathbf{r}_{e}, X_{i}, t\right)$. With coordinates transform (\ref{5})
\begin{eqnarray}
&\displaystyle 
\hat{x}=x_{i}-\frac{e}{m_{i}}\frac{\cos \omega_{0}t}{\omega_{0}\left(\omega^{2}_{0}-\omega^{2}_{ci}\right)}\left(\omega_{0}E_{0x}\left(X_{i}\right)-
\omega_{ci}E_{0y}\left(X_{i}\right)\right)
\label{18}
\end{eqnarray}
and 
\begin{eqnarray}
&\displaystyle 
\hat{y}=y_{i}-\frac{e}{m_{i}}\frac{\sin\omega_{0}t}{\omega_{0}\left(\omega^{2}_{0}-\omega^{2}_{ci}\right)}
\left(\omega_{ci}E_{0x}\left(X_{i}\right)-\omega_{0}E_{0y}\left(X_{i}\right)\right),
\label{19}
\end{eqnarray}
for the ions and with similar transforms for the electrons 
we obtain the relation for the Fourier transform in the ion frame of the electron density perturbation,
\begin{eqnarray}
& \displaystyle 
\int d\mathbf{r}_{i}\delta n_{e}\left(\mathbf{r}_{e}, t\right)e^{-i\mathbf{k}_{i}\mathbf{r}_{i}}=\sum\limits_{m=-\infty}^{\infty}
J_{m}\left(a_{ei}\right)
\nonumber\\
&\displaystyle 
\times
e^{im\left(\omega_{0}t+\delta\right)}\delta n_{e}^{(e)}\left(\mathbf{k}_{i}, t\right),
\label{20}
\end{eqnarray}
and the identity
\begin{eqnarray}
& \displaystyle 
\varphi^{(e)}_{e}\left(\mathbf{k}_{e}, t\right)=e^{-ia_{ei}\sin\left(\omega_{0}t+\delta_{i}\right)}\varphi_{i}\left(\mathbf{k}_{i}, t\right)
\nonumber
\\ 
&\displaystyle
=\sum\limits_{m=-\infty}^{\infty}
J_{p}\left(a_{ei}\right)e^{ip\left(\omega_{0}t+\delta\right)}\varphi_{i}\left(\mathbf{k}_{i}, t\right),
\label{21}
\end{eqnarray}
which determines the relation of the Fourier transform $\varphi^{(e)}_{e}\left(\mathbf{k}_{e}, t\right)$ of the potential $\varphi\left(\mathbf{r}_{e}, t
\right)$ determined in the electron frame with potential $\varphi_{i}\left(\mathbf{k}_{i}, t\right)$. This relation is the manifestation of the Doppler 
effect in the case of two oscillating frames of references. It follows from Eq. (\ref{21}) that an elementary sine
disturbance with frequency $\omega$ in the ion frame of reference is recognized in the electron frame connected with the electron component 
as an infinite set of disturbances with frequencies $\omega +p\omega_{0}$.
The parameters $a_{ei}$ and $\delta$ in Eqs. (\ref{20}) and (\ref{21}) for electric field (\ref{1}) are
\begin{widetext}
\begin{eqnarray}
& \displaystyle 
a_{ei}= a_{ei}\left(X_{i}\right)=\left\lbrace\left[\sum\limits_{\alpha=i, e}\frac{e_{\alpha}k_{iy}}{2m_{\alpha}\omega_{0}}\left(\frac{E_{0x}-E_{0y}}
{\omega_{0}-\omega_{c\alpha}} -\frac{E_{0x}+E_{0y}}{\omega_{0}+\omega_{c\alpha}}\right)  \right]^{2}\right.
\nonumber
\\ 
&\displaystyle
\left.+\left[\sum\limits_{\alpha=i, e}\frac{e_{\alpha}k_{ix}}{2m_{\alpha}\omega_{0}}\left(\frac{E_{0x}+E_{0y}}
{\omega_{0}+\omega_{c\alpha}} +\frac{E_{0x}-E_{0y}}{\omega_{0}-\omega_{c\alpha}}\right) \right]^{2}  \right\rbrace ^{1/2} = |\mathbf{k}_{i}\xi_{ie}|,
\label{22}
\end{eqnarray}
\end{widetext}
where $E_{0x}=E_{0x}\left(X_{i}\right)$ and $E_{0y}=E_{0y}\left(X_{i}\right)$ and $\xi_{ie}$ is the amplitude of the displacements of electrons 
relative ions in FW, and 
\begin{eqnarray}
& \displaystyle 
\tan \delta\left(X_{i}\right)=\frac{\sum\limits_{\alpha=i, e}\frac{e_{\alpha}k_{iy}}{m_{\alpha}}\left(\frac{E_{0x}+E_{0y}}{\omega_{0}+
\omega_{c\alpha}}+\frac{E_{0x}-E_{0y}}
{\omega_{0}-\omega_{c\alpha}}\right)  
}{\sum\limits_{\alpha=i, e}\frac{e_{\alpha}k_{ix}}{m_{\alpha}}\left(\frac{E_{0x}+E_{0y}}
{\omega_{0}+\omega_{c\alpha}} -\frac{E_{0x}-E_{0y}}{\omega_{0}-\omega_{c\alpha}}\right)}.
\label{23}
\end{eqnarray}
Using Eqs. (\ref{20}) and (\ref{21}) in the Fourier transformed Poisson 
equation (\ref{17}) for the potential $\varphi_{i}\left(\mathbf{k}_{i}, t\right)$, we derive the basic equation for 
$\varphi_{i}\left(\mathbf{k}_{i}, \omega\right)$,
\begin{eqnarray}
&\displaystyle \varepsilon\left(\mathbf{k}_{i}, \omega\right)\varphi_{i}\left(\mathbf{k}_{i}, X_{i}, \omega\right)
+\sum\limits_{q\neq 0}\sum\limits_{m=-\infty}^{\infty}J_{m}\left(a_{ei}\right)J_{m+q}\left(a_{ei} \right)
\nonumber
\\ 
&\displaystyle
\times e^{iq\delta}\varepsilon_{e}\left(\mathbf{k}_{i}, \omega-m\omega_{0}\right)
\varphi_{i}\left(\mathbf{k}_{i}, X_{i}, \omega+q\omega_{0}\right)=0,
\label{24}
\end{eqnarray}
where
\begin{eqnarray}
&\displaystyle \varepsilon\left(\mathbf{k}_{i}, X_{i}, \omega\right)=1+\varepsilon_{i}\left(\mathbf{k}_{i}, X_{i}, \hat{\omega}\right)
\nonumber
\\ 
&\displaystyle
+ \sum\limits_{m=-\infty}^{\infty}
J^{2}_{m}\left(a_{ei}\right)\varepsilon_{e}\left(\mathbf{k}_{i}, X_{i}, \omega-m\omega_{0}\right).
\label{25}
\end{eqnarray}
In Eq. (\ref{25}), the function $\varepsilon_{i}\left(\mathbf{k}_{i}, X_{i}, \hat{\omega}\right)$ is the renormalized dielectric 
permittivity of ions \cite{Mikhailenko3}, which accounts for the nonlinear effect of the scattering of ions by the electrostatic IC turbulence; 
the function  $\varepsilon_{e}\left(\mathbf{k}_{i}, X_{i}, \omega\right)$ is the linear permittivity of electrons. 
For the inhomogeneous plasma with 
the  Maxwellian distribution (\ref{14}) for ions and electrons with inhomogeneous density and temperature  these functions are
\begin{eqnarray}
&\displaystyle \varepsilon_{i}\left(\mathbf{k}_{i}, \hat{\omega}\right)=\frac{1}{k_{i}^2\lambda _{Di }^2}\left[1 + i\sqrt {\pi}
\left(z_{i0}-\chi_{i}\left(1-\frac{1}{2}\eta_{i}\right)\right)\right.
\nonumber
\\ 
&\displaystyle
\times \sum\limits_{n = - \infty }^{\infty}W\left(z_{in}\right)A_{in}\Big] 
\nonumber
\\ 
&\displaystyle
-\eta_{i}\chi_{i}\sum\limits_{n = - \infty }^{\infty}z_{in}\left(1+i\sqrt{\pi}z_{in}W\left(z_{in}\right)\right)A_{in}
\nonumber
\\ 
&\displaystyle +\eta_{i}\chi_{i}\sum\limits_{n = - \infty}^{\infty}i\sqrt{\pi}W\left(z_{in}\right)k^{2}_{i\bot}\rho^{2}_{i}
\left(A_{in}-\hat{A}_{in}\right),
\label{26}
\end{eqnarray}
\begin{eqnarray}
&\displaystyle 
\varepsilon_{e}\left(\mathbf{k}_{i}, \omega-m\omega_{0}\right)
=\frac{1}{k_{i}^2\lambda_{De}^2}\left(1+i\sqrt{\pi}\left(z_{e}\right.\right.
\nonumber
\\ 
&\displaystyle
\left.\left.-\chi_{e}\left(1-\frac{1}{2}\eta_{e}\right)-m\zeta_{e}\right)W\left(z_{e}-m\zeta_{e}\right)\right).
\label{27}
\end{eqnarray}
In Eqs. (\ref{26}), (\ref{27}), $\lambda_{Di(e)}$ is the ion (electron) Debye length, $A_{in}=I_{n}\left(k^{2}_{i\bot}\rho^{2}_{i}\right)\exp\left(-k^{2}_{i
\bot}\rho^{2}_{i}\right)$, $\hat{A}_{in}=I'_{n}\left(k^{2}_{i\bot}\rho^{2}_{i}\right)\exp\left(-k^{2}_{i\bot}\rho^{2}_{i}\right)$, $I_{n}$ is the modified 
Bessel function of order $n$, $\rho_{i}= v_{Ti}/\omega_{ci}$ is the ion thermal Larmor radius, $W\left(z\right)=e^{-z^{2}}\left(1 +\left(2i/ \sqrt {\pi } 
\right)\int\limits_{0}^{z} e^{t^{2}}dt \right)$ is the complex error function, $z_{in} =\left(\hat{\omega}-n\omega_{ci}\right)/\sqrt{2}k_{iz}v_{Ti}$, 
$\chi_{i} = k_{iy}v_{di}/\sqrt{2}k_{iz}v_{Ti}$, $\eta_{i}=d \ln T_{i}/d \ln n_{0i}$, $v_{di,e}= \left( cT_{i,e}/eB_{0}\right) \left( 
d\ln n_{i}/dx\right) $ is the ion (electron) diamagnetic velocity, 
$z_{e} = \omega/\sqrt{2}k_{iz}v_{Te}$, $\chi_{e} = k_{iy}v_{de}/\sqrt{2}k_{iz}v_{Te}$, and $\zeta_{e} = \omega_{0}/\sqrt{2}k_{iz}v_{Te}$. 
The solution to Eq. (\ref{24}) is the renormalized frequency $\hat{\omega}\left(\mathbf{k}_{i}\right)=\omega\left(\mathbf{k}_{i}\right)+iC_{i}
\left(\mathbf{k}_{i}\right)$, where $C_{i}\left(\mathbf{k}_{i}\right)$ is determined by the equation\cite{Mikhailenko3}, 
\begin{eqnarray}
& \displaystyle
\left\langle e^{-i\mathbf{k}_{i\bot}\left(\delta\mathbf{r}\left(t\right)-\delta\mathbf{r}\left(t_{1}\right)\right)}\right\rangle
\nonumber
\\
&\displaystyle 
\simeq e^{-\frac{1}{2}\left\langle\left(\mathbf{k}_{i\bot}\delta\mathbf{r}
\left(t-t_{1}\right)\right)^{2}\right\rangle}=e^{-C_{i}\left(t-t_{1}\right)},
\label{28}
\end{eqnarray}
where  $\mathbf{k}_{i}\delta\mathbf{r}_{i}\left(t\right)$ is the nonlinear phase shift, 
resulted from the perturbations of the ions orbits by their interaction with IC turbulence. The coefficient $C_{i}$ is equal to\cite{Mikhailenko3} 
\begin{eqnarray}
& \displaystyle C_{i}=\frac{e^{2}}{2m^{2}_{i}}Re \sum\limits_{n=-\infty}^{\infty}\int d\mathbf{k}_{i1}|
\varphi\left(\mathbf{k}_{i1\bot}\right)|^{2}
\mathcal{F}_{i}\left( k_{i\bot},k_{i1\bot}\right)
\nonumber
\\
&\displaystyle 
\times
e^{-k_{i\bot}^{2}\rho^{2}_{i}}\sqrt{\frac{\pi}{2}}\frac{1}{k_{iz}v_{Ti}}
W\left(\frac{\omega-n_{1}\omega_{ci}}{\sqrt{2}k_{iz}v_{Ti}}\right),
\label{29}
\end{eqnarray}
with
\begin{eqnarray}
& \displaystyle \mathcal{F}_{i}\left( k_{i\bot},k_{i1\bot}\right)=\frac{2}{\omega^{2}_{ci}}
\left(k_{ix}k_{i1y}-k_{iy}k_{i1x}\right)^{2}
I_{n}\left(k^{2}_{i\bot}\rho^{2}_{i}\right)
\nonumber\\ 
& \displaystyle
+\frac{1}{2}\frac{k^{2}_{i\bot}k^{2}_{i1\bot}}{\omega^{2}_{ci}}\Big(I_{n+1}\left(k^{2}_{i\bot}
\rho^{2}_{i}\right)
+I_{n-1}\left(k^{2}_{i\bot}\rho^{2}_{i}\right)\Big).
\label{30}
\end{eqnarray}

Equation (\ref{24}) is in fact the infinite system of equations for the potential $\varphi_{i}\left(\mathbf{k}_{i}, \omega-q_{0}\omega_{0}\right)$
coupled with $\varphi_{i}\left(\mathbf{k}_{i}, \omega-q(\neq q_{0})\omega_{0}\right)$. The detailed numerical solution of Eq. (\ref{24})  was performed in 
Ref. \cite{Mikhailenko3} for the model of three interacting modes $\varphi_{i}\left(\mathbf{k}
_{i}, \omega\right)$, $\varphi_{i}\left(\mathbf{k}_{i}, \omega+m\omega_{0}\right)$ and $\varphi_{i}\left(\mathbf{k}_{i}, \omega-m\omega_{0}\right)$. 
This model involves IC kinetic parametric instability considered in Ref.[12], 
the quasimode decay instability considered in Ref.[13] which develop due to the coupling $\varphi_{i}\left(\mathbf{k}_{i}, \omega\right)$ 
with other IC mode $\varphi_{i}\left(\mathbf{k}_{i}, \omega+m_{0}\omega_{0}\right)$, and 
possible instabilities which are caused by the three modes coupling. For the detailed results of the numerical solution of Eq. (\ref{24}) we refer to our paper\cite{Mikhailenko3}. The general conclusion, which follows from this analysis, is that the inverse electron Landau 
damping is the decisive process in the development of the parametric instability with maximum growth rate. The most unstable IC perturbations have wavelength 
$k_{i\bot}\rho_{i}\approx k_{iy}\rho_{i}\approx  1.5$. The spectral intensity $|\varphi_{i}\left(\mathbf{k}_{i}\right)|^{2}$ of the IC parametric turbulence 
at the saturation state, which establishes due to the nonlinear scattering of ions by the electric field of the IC turbulence, is determined from 
the nonlinear integral equation\cite{Mikhailenko3, Dum}
\begin{eqnarray}
&\displaystyle
\gamma\left(\mathbf{k}_{i}, X_{i}\right)=C_{i}\left(\mathbf{k}_{i}, X_{i}\right).
\label{31}
\end{eqnarray}
It follows from Eq. ({31}) that the local spectral intensity at the steady 
state on the microscale spatial scales depends on the spatial mesoscale variable $X_{i}$, which enters to the solution for 
$|\varphi_{i}\left(\mathbf{k}_{i}, X_{i}\right)|^{2}$ as a parameter, included in $\gamma\left(\mathbf{k}_{i}, X_{i}\right)$ and 
$C_{i}\left(\mathbf{k}_{i}, X_{i}\right)$. It is obvious that the exact dependence of $|\varphi_{i}\left(\mathbf{k}_{i}, X_{i}\right)|^{2}$ on $X_{i}$
for the steady state for the IC microturbulence may be determined only numerically for the known distribution of the plasma density, of the temperature of 
the plasma components and of the spatial 
distribution of the FW electric field, which determines the spatial distribution of the parameters $a_{ei}\left(X_{i}\right)$ and $\delta\left(X_{i}\right)$ 
in Eq. {(24}). 

The simple estimate for the solution to Eq. ({31}) for $|\varphi_{i}\left(\mathbf{k}_{i}, X_{i}\right)|^{2}$ may be derived 
employing the mean value theorem for the integral over $\mathbf{k}$ in Eq. (\ref{29}) with assumption that the spectrum of 
fluctuations $|\varphi_{i}\left(\mathbf{k}_{i}, X_{i}\right)|^{2}$ is peaked near the linearly most unstable 
wavenumber, $\mathbf{k}_{0}$.  With this approximation the integral equation for $|\varphi_{i}\left(\mathbf{k}_{i}, X_{i}\right)|^{2}$ 
transforms into the algebraic relation for the root-mean-square (rms) magnitude, 
\begin{eqnarray}
&\displaystyle
\widetilde{\varphi}\left(X_{i}\right)=\left(\int\left|\varphi\left(X_{i}, \mathbf{k}_{1}\right)\right|^{2}d\mathbf{k}_{1}\right)^{1/2},
\label{32}
\end{eqnarray}
of the electrostatic potential. For $|z_{in}|\gg 1$ that  gives for $\widetilde{\varphi}\left(X_{i}\right)$ the estimate\cite{Mikhailenko2,Mikhailenko3}
\begin{eqnarray}
&\displaystyle
\frac{e\widetilde{\varphi_{i}}}{T_{i}}\sim\left(\frac{1}{k_{0}\rho_{i}}\right)^{5/2}\sim \left(\frac{U}{v_{Ti}}\right)^{5/2}.
\label{33}
\end{eqnarray}
It follows from Eq. (\ref{33}) that the condition of the comparable values of the current velocity amplitude $U$ and of the ion thermal velocity 
$v_{Ti}$  is in fact the condition for the 'physical' threshold at which the level of the IC parametric turbulence becomes substantional; 
this level reduces sharply in the top of the 
pedestal and in a bulk of plasma where ion thermal velocity usually is much larger than the current velocity $U$. This turbulence may have high 
level in the pedestal and near-SOL low temperature plasma   near the FW antenna.  For the numerical data in 
Ref.\cite{Bertelli, Bertelli1} $U\sim cE_{0x}/B_{0}\approx 4,5 \cdot 10^{6}$\,cm/s and 
$v_{Ti}\approx 4,4 \cdot 10 ^{6}$cm/s $\sim U$ for deuterium ion with temperature $T_{i}=40$ eV. For these data the maximum growth rate 
have the IC perturbations with $k_{i\bot}\rho_{i}\sim 1$. It was found in Ref.\cite{Mikhailenko3} that the energy density 
$W=\int W\left(\mathbf{k}_{i}\right)d\mathbf{k}_{i}$, where
\begin{eqnarray*}
&\displaystyle
W\left(\mathbf{k}_{i}\right)=k^{2}_{i}\omega\left(\mathbf{k}_{i}\right)
\left| \varphi\left(\mathbf{k}_{i}\right)\right| ^{2}
\frac{\partial \varepsilon_{i}}
{\partial \omega\left(\mathbf{k}_{i}\right)},
\end{eqnarray*}
for these IC perturbations in the saturated state is $W/n_{0i}T_{i}\sim 1$.

The effect of the inhomogeneity of the plasma density included in Eqs. (\ref{26}) and (\ref{27}) by the frequencies $\omega_{di,e}=k_{y}v_{di, e}$
does not affect the frequency $\omega\left(\mathbf{k}_{i}\right)\sim n\omega_{ci}$ and the growth rate of the IC  parametric instabilities  when $n
\omega_{ci}\gg k_{y}v_{di,e}$. This occurs for the perturbations for which $k_{y}\rho_{i}< L_{n}/\rho_{i}$ 
and the estimate for the IC potential is given by Eq. (\ref{33}).

The ion density inhomogeneity is substantial for the short wavelength perturbations with
\begin{eqnarray}
&\displaystyle
k_{iy}\rho_{i}>\frac{L_{n}}{\rho_{i}}\gg 1,
\label{34}
\end{eqnarray}
However, the level of the IC turbulence for this part of the IC turbulence spectrum, which is established 
due to the scattering of IC perturbation by the IC turbulence, is negligibly small, 
\begin{eqnarray}
&\displaystyle
\frac{e\widetilde{\varphi}_{i}}{T_{i}}\sim\left(\frac{1}{k_{0}\rho_{i}}\right)^{5/2}\lesssim \left(\frac{\rho_{i}}{L_{n}}\right)^{5/2} \ll 1.
\label{35}
\end{eqnarray}
Therefore the effect of the drift-cyclotron instability, which may be developed in this part of the IC spectrum, on the level of IC turbulence 
of the inhomogeneous pedestal plasma is negligible small. 

The temporal evolution of the ion distribution function $F_{0i}\left(\mathbf{v}_{i}, X_{i}, t\right)$ under the action of the microscale IC 
turbulence is determined by the known quasilinear equation, which in our case has a form
\begin{eqnarray}
&\displaystyle 
\frac{\partial F_{i0}\left(\mathbf{v}_{i}, X_{i}, t\right)}{\partial t}
=\frac{e_{i}}{m_{i}}\left\langle \frac{\partial\varphi\left(\mathbf{r}_{i}, X_{i}, t\right)}{\partial
\mathbf{r}_{i}}\frac{\partial f_{i}\left(\mathbf{v}_{i}, \mathbf{r}_{i}, X_{i}, t\right)}{\partial \mathbf{v}_{i}}\right\rangle,
\label{36}
\end{eqnarray}
derived from Eq. (\ref{13}). The angle brackets $\left\langle ...\right\rangle $ indicate the averaging 
of the expression in it over the initial phases of the microscale IC perturbations. The particular forms of Eq. (\ref{36}) were derived in 
Ref.\cite{Mikhailenko2} for the IC turbulence powered 
by the IC kinetic parametric instability, and in Ref.\cite{Mikhailenko3} for the IC turbulence powered by the quasimode decay instability.
The quasilinear equation (\ref{36}) is the basic equation for the derivation, as the corresponding moments of this equation, the 
equations which govern the temporal evolution of the ion density and the ion thermal energy. By multiplying Eq. (\ref{36}) on $m_{i}v_{i\bot}^{2}/2$ and 
integrating it over velocities, we found the equation 
\begin{eqnarray}
&\displaystyle \frac{\partial n_{0i}T_{i\bot}}{\partial t}= Q_{i}\left(X_{i}, t\right),
\label{37}
\end{eqnarray} 
which governs the temporal growth at fixed radial position $X_{i}$ of the averaged density of the perpendicular thermal energy of 
ions, $n_{0i}\left(X_{i}, t\right)T_{i\bot}\left(X_{i}, t\right)$, determined as
\begin{eqnarray}
&\displaystyle n_{0i}\left(X_{i}, t\right)T_{i\bot}\left(X_{i}, t\right)
\nonumber\\ 
& \displaystyle
=\frac{1}{2}\int d\mathbf{v}_{i}
F_{i0}\left(\mathbf{v}_{i}, X_{i}, t\right)m_{i}v^{2}_{i\bot}.
\label{38}
\end{eqnarray}
The function $Q_{i}\left(X_{i}, t\right)$ in Eq. (\ref{37}) is determined by the relation
\begin{eqnarray}
&\displaystyle Q_{i}\left(X_{i}, t\right)=\frac{1}{2}\int d\mathbf{v}_{i}m_{i}v^{2}_{i\bot}
\nonumber\\ 
&\displaystyle
\times\left\langle \frac{e_{i}}{m_{i}}\frac{\partial\varphi\left(\mathbf{r}_{i},t\right)}{\partial
\mathbf{r}_{i}}\frac{\partial f_{i}\left(\mathbf{v}_{i}, \mathbf{r}_{i}, X_{i}, t\right)}{\partial \mathbf{v}_{i}}\right\rangle.
\label{39}
\end{eqnarray}  
For the IC turbulence developed in the near antenna region due to the IC kinetic parametric instability\cite{Mikhailenko3}, the growth of the perpendicular 
thermal energy of ions resulted from the scattering of the nonresonant ions by the IC parametric turbulence was estimated in 
Refs. \cite{Mikhailenko1, Mikhailenko2} as
\begin{eqnarray}
&\displaystyle n_{0i}\frac{\partial T_{i\bot}}{\partial t}\sim \gamma \frac{W}{n_{0i}
T_{i\bot}}n_{0i}T_{i\bot},
\label{40}
\end{eqnarray} 
This estimate was derived under the assumption that the spectrum of fluctuations is peaked near the linearly most unstable wavenumber with the 
growth rate of the instability $\gamma$; $W$ is the energy density of the IC turbulence in the saturated state. 

The similar relation for $T_{i\bot}$ growth holds when IC turbulence is powered by the IC quasimode decay 
instability\cite{Mikhailenko3} 
and the interactions of ions with IC turbulence occurs under conditions of the IC resonance\cite{Mikhailenko3}.

The  solution of Eq. (\ref{24}) for the complex frequency $\omega\left(\mathbf{k}_{i}, X_{i}\right)$  of the possible parametric IC 
instabilities, the solution of Eq. (\ref{31}) for the spectral intensity $|\varphi_{i}\left(\mathbf{k}_{i}, X_{i}\right)|^{2}$ 
of the electrostatic potential in the steady state, and the solutions of Eq. (\ref{40}) for the ion temperature all contain
the mesoscale spatial variable $X_{i}$ which originate from the equilibrium distribution function (\ref{14}) of inhomogeneous plasma and inhomogeneous 
FW field (\ref{1}). It is obvious, that all these solutions with predicted plasma and FW inhomogeneities may be derived only numerically. The  local 
approximation developed above simplifies partially the solution of this problem.

\section{Mesoscale plasma flows driven by the inhomogeneous microturbulence}\label{sec4}

Equation (\ref{31}) reveals that the microscale IC parametric turbulence of the inhomogeneous plasma driven by the inhomogeneous FW field is spatially
inhomogeneous along coordinate $X_{i}$ in the saturation state. The electric field $\tilde{\mathbf{E}}$ of this turbulence is determined in the form
\begin{eqnarray}
&\displaystyle 
\tilde{\mathbf{E}}\left(\mathbf{r}_{i}, X_{i}, t\right)=\int d\mathbf{k}\tilde{\mathbf{E}}\left(\mathbf{k}, X_{i}\right)
\nonumber\\ 
& \displaystyle
\times e^{-i\omega\left(\mathbf{k}, X_{i}\right)t+ i\mathbf{k}\mathbf{r}_{i}+i\theta\left(\mathbf{k}\right)}
\nonumber\\ 
& \displaystyle
=-i\int d\mathbf{k}\mathbf{k}\varphi\left(\mathbf{k}, X_{i}\right)
\nonumber\\ 
& \displaystyle
\times e^{-i\omega\left(\mathbf{k}, X_{i}\right)t+ i\mathbf{k}\mathbf{r}_{i}+i\theta\left(\mathbf{k}\right)},
\label{41}
\end{eqnarray}
where the integration over $\mathbf{k}$ is performed over wave numbers of the linearly unstable IC perturbations, and 
$\theta\left(\mathbf{k}\right)$ is their initial phase.  
In this equation, electric field $\tilde{\mathbf{E}}\left(\mathbf{k}, X_{i}\right)=-i\mathbf{k}\varphi\left(\mathbf{k}, X_{i}\right)$ is determined  
by Eq. (31) as the electric field of the IC turbulence in the saturation state.  The spatial inhomogeneity scale length of the turbulent electric field 
is much larger than the wavelengths of the IC parametric turbulence perturbations, but is commensurable  with the density inhomogeneity 
scale length and of the FW electric field scale length in the near antenna region. In this section, we consider the average 
effect of the spatial inhomogeneity of IC turbulence on the mesoscale temporal evolution of the plasma in the FW near-antenna region.  
For this goal, we perform the transformation of the $\mathbf{r}_{i}$ and $\mathbf{v}_{i}$ variables in Eq. (\ref{13}) for 
$F_{i}\left( \mathbf{v}_{i}, \mathbf{r}_{i}, X_{i}, t\right)$ to new spatial $\mathbf{\tilde{r}}_{i}$ and velocity $\mathbf{\tilde{v}}_{i}$ coordinates, 
in which the ion thermal motion and the ion motion in the electric field (\ref{41}) of the IC parametric turbulence are separated. 
These new coordinates are determined by the relations 
\begin{eqnarray}
&\displaystyle
\mathbf{v}_{i}=\mathbf{\tilde{v}}_{i}+\mathbf{\tilde{U}}_{i}\left(\mathbf{\tilde{r}}_{i}, X_{i}, t\right) 
\label{42}
\end{eqnarray}
and 
\begin{eqnarray}
&\displaystyle
\mathbf{r}_{i}=\mathbf{\tilde{r}}_{i}+\mathbf{\tilde{R}}_{i}\left(\mathbf{\tilde{r}}_{i}, t\right)= \mathbf{\tilde{r}}_{i}+\int\limits^{t}_{t_{0}}
\mathbf{\tilde{U}}_{i}\left(\mathbf{\tilde{r}}_{i}, X_{i}, t_{1}\right)dt_{1}, 
\label{43}
\end{eqnarray}
or by their inverse relations
\begin{eqnarray}
&\displaystyle
\mathbf{\tilde{v}}_{i}=\mathbf{v}_{i}-\mathbf{\tilde{V}}_{i}\left(\mathbf{r}_{i}, X_{i}, t\right) 
\label{44}
\end{eqnarray}
and 
\begin{eqnarray}
&\displaystyle
\mathbf{\tilde{r}}_{i}=\mathbf{r}_{i}-\int\limits^{t}_{t_{0}}\mathbf{\tilde{V}}_{i}\left(\mathbf{r}_{i}, X_{i}, t_{1}\right)dt_{1}. 
\label{45}
\end{eqnarray}
The velocity $\mathbf{\tilde{V}}_{i}\left(\mathbf{r}, X_{i}, t\right)$ is determined by the equation 
\begin{eqnarray}
&\displaystyle 
\frac{\partial \mathbf{\tilde{V}}_{i}}{\partial t}+\mathbf{\tilde{V}}_{i}\frac{\partial \tilde{\mathbf{V}}_{i}}{\partial 
\mathbf{r}_{i}}
\nonumber\\ 
& \displaystyle
=\frac{e_{i}}{m_{i}}\left(\mathbf{\tilde{E}}\left(\mathbf{r}_{i}, X_{i}, t\right)
+\frac{1}{c}\left[\tilde{\mathbf{V}}_{i}\times\mathbf{B}_{0}\right]\right).
\label{46}
\end{eqnarray}
In this equation, electric field $\tilde{\mathbf{E}}\left(\mathbf{k}, X_{i}\right)=-i\mathbf{k}\varphi\left(\mathbf{k}, X_{i}\right)$ is determined in Eq. 
(41) as the electric field of the IC turbulence in the saturation state by Eq. (31).
The solutions to Eq. (\ref{46}) with initial value $\mathbf{\tilde{V}}_{i}\left(\mathbf{r}_{i}, t_{0} \right)=0$ are
\begin{eqnarray}
&\displaystyle 
\tilde{V}_{ix}\left(\mathbf{r}_{i}, X_{i}, t \right) = \frac{e}{m_{i}}\int\limits ^{t}_{t_{0}}dt_{1}\left[\tilde{E}_{x}
\left(\mathbf{r}_{i}, X_{i}, t_{1}\right)\cos \omega_{ci}\left(t-t_{1}\right)\right.
\nonumber 
\\ &\displaystyle
\left. + \tilde{E}_{y}\left(\mathbf{r}_{i}, X_{i}, t_{1}\right)\sin \omega_{ci}\left(t-t_{1}\right) \right], 
\label{47}
\end{eqnarray}
and
\begin{eqnarray}
&\displaystyle 
\tilde{V}_{iy}\left(\mathbf{r}_{i}, X_{i}, t \right) = \frac{e}{m_{i}}\int\limits ^{t}_{t_{0}}dt_{1}\left[-\tilde{E}_{x}
\left(\mathbf{r}_{i}, X_{i}, t_{1}\right)\sin\omega_{ci}\left(t-t_{1}\right)\right.
\nonumber 
\\ &\displaystyle
\left. + \tilde{E}_{y}\left(\mathbf{r}_{i}, X_{i}, t_{1}\right)\cos \omega_{ci}\left(t-t_{1}\right) \right]. 
\label{48}
\end{eqnarray}
Using solutions (\ref{47}), (\ref{48}) for $\mathbf{\tilde{V}}_{i}$ in 
the Vlasov equation (\ref{13}) and averaging this equation over the initial phases of the IC perturbations and over the time 
$t\gg \omega^{-1}\left(\mathbf{k}\right) \sim \omega^{-1}_{ci}$, we derive the kinetic equation which determines the spatial and temporal 
evolution of the ion distribution function ${\textit {F}}_{i}$ on the intermediate spatial scales commensurable with spatial scales of the 
nonuniformities of the equilibrium plasma and FW. This equation has a form
\begin{eqnarray}
&\displaystyle 
\frac{\partial F_{i}}{\partial t}
+\left(v_{ix}-\bar{U}_{ix}\left(X_{i}\right)\right)\frac{\partial F_{i}}{\partial X_{i}}
\nonumber\\ 
& \displaystyle
+\left(v_{iy}-\bar{U}_{iy}\left(X_{i}\right)\right)\frac{\partial F_{i}}{\partial Y_{i}}=0.
\label{49}
\end{eqnarray}
The initial condition for this equation is the equilibrium ion distribution function $F_{i0}\left(\mathbf{v}_{i}, X_{i}\right)$. 
In our case it is determined by the Maxwellian distribution (\ref{14}) for the pedestal plasma with possible quasilinear temporal evolution of the ion 
density and temperature, determined by Eq. (\ref{36}), when this evolution occurs at the time before the onset of the development of the mesoscale evolution. 
In Eq. (\ref{49}), $\bar{U}_{ix}\left(X_{i}\right)$ and $\bar{U}_{iy}\left(X_{i}\right)$ are the velocities of the nonlinear convective flow, determined by 
the relations
\begin{eqnarray}
&\displaystyle 
\bar{U}_{ix}\left(X_{i}\right) = \left\langle \tilde{V}_{ix}\left(\mathbf{\tilde{r}}_{i}, X_{i}, t \right)\frac{\partial}{\partial X_{i}} \int\limits^{t}
_{t_{0}}\tilde{V}_{ix}\left(\mathbf{\tilde{r}}_{i}, X_{i}, t_{1} \right)dt_{1}\right\rangle, 
\label{50}
\end{eqnarray}
and
\begin{eqnarray}
&\displaystyle 
\bar{U}_{iy}\left(X_{i}\right) = \left\langle \tilde{V}_{ix}\left(\mathbf{\tilde{r}}_{i}, X_{i}, t \right)\frac{\partial}{\partial X_{i}} \int\limits^{t}
_{t_{0}}\tilde{V}_{iy}\left(\mathbf{\tilde{r}}_{i}, X_{i}, t_{1} \right)dt_{1}\right\rangle,
\label{51}
\end{eqnarray}
where velocities $\tilde{V}_{ix}\left(\mathbf{\tilde{r}}_{i}, X_{i}, t_{1} \right)$ and $\tilde{V}_{iy}\left( \mathbf{\tilde{r}}_{i}, X_{i}, t_{1} \right)$
are determined above by Eqs. (\ref{47}) and (\ref{48}). The angle brackets $\left\langle ...\right\rangle $ indicate the averaging 
of the expression in it over the initial phases of the IC perturbations. The averaged over time $t\gg \omega^{-1}\left(\mathbf{k}\right) \sim \omega^{-1}
_{ci}$ values of $\bar{U}_{ix}\left(X_{i}\right)$ and $\bar{U}_{iy}\left(X_{i}\right)$ are uniform in time and are equal to
\begin{widetext}
\begin{eqnarray}
&\displaystyle 
\bar{U}_{ix}\left(X_{i}\right)= \frac{1}{2\omega_{ci}}\frac{e^{2}}{m^{2}_{i}}\int d\mathbf{k}\left[a_{i1}\left(\mathbf{k}\right) \tilde{E}_{x}\left( 
\mathbf{k}, X_{i}\right)
\frac{\partial}{\partial X_{i}}\left(\tilde{E}^{\ast}_{y}\left(\mathbf{k}, X_{i}\right)\right)\right.
\nonumber\\ 
& \displaystyle
\left.+a_{i2}\left(\mathbf{k}\right)\tilde{E}_{y}\left(\mathbf{k}, X_{i}\right)\frac{\partial}{\partial X_{i}}
\left(\tilde{E}^{\ast}_{x}\left(\mathbf{k}, X_{i}\right)\right)\right]
\nonumber\\ 
& \displaystyle
= \frac{1}{4\omega_{ci}}\frac{e^{2}}{m^{2}_{i}}\int d\mathbf{k}k_{x}k_{y}\left(a_{i1}\left(\mathbf{k}\right) +a_{i2}\left(\mathbf{k}\right) \right)
\frac{\partial}{\partial X_{i}}\left|\varphi\left(\mathbf{k}, X_{i}\right)\right|^{2},
\label{52}
\end{eqnarray}
and
\begin{eqnarray}
&\displaystyle 
\bar{U}_{iy}\left(X_{i}\right)= -\frac{1}{4\omega_{ci}}\frac{e^{2}}{m^{2}_{i}}\int d\mathbf{k}\left[ a_{i1}\left(\mathbf{k}\right)\frac{\partial}{\partial 
X_{i}}\left|\tilde{E}_{x}\left(\mathbf{k}, X_{i}\right)\right|^{2}
-a_{i2}\left(\mathbf{k}\right)\frac{\partial}{\partial X_{i}}
\left|\tilde{E}_{y}\left(\mathbf{k}, X_{i}\right)\right|^{2}\right]
\nonumber\\ 
& \displaystyle
= -\frac{1}{4\omega_{ci}}\frac{e^{2}}{m^{2}_{i}}\int d\mathbf{k}\left(k^{2}_{x}a_{i1}\left(\mathbf{k}\right)-k^{2}_{y}a_{i2}\left(\mathbf{k}\right) 
\right)\frac{\partial}{\partial X_{i}}\left|\varphi\left(\mathbf{k}, X_{i}\right)\right|^{2},
\label{53}
\end{eqnarray}
where the asterisk in Eq. (\ref{52}) implies the operation of complex conjugate. The coefficients $a_{i1}\left(\mathbf{k}\right)$ 
and $a_{i2}\left(\mathbf{k}\right)$ are determined as
\begin{eqnarray}
&\displaystyle a_{i1}\left(\mathbf{k}\right)=\left[\frac{\omega_{ci}}{\omega\left(\mathbf{k}\right)\left(\omega_{ci}+\omega\left(\mathbf{k}\right)
\right)^{2}} +\frac{\omega_{ci}}
{\omega\left(\mathbf{k}
\right)\left(\omega_{ci}-\omega\left(\mathbf{k}\right)\right)^{2}}+
\frac{1}{\left(\omega^{2}_{ci}-\omega^{2}\left(\mathbf{k}\right)\right)}\right], 
\label{54}
\end{eqnarray}
and
\begin{eqnarray}
&\displaystyle a_{i2}\left(\mathbf{k}\right)=\left[\frac{1}{\left(\omega_{ci}+\omega\left(\mathbf{k}\right)\right)^{2}} +\frac{1}{\left(\omega_{ci}-\omega
\left(\mathbf{k}\right)
\right)^{2}}+\frac{1}{\left(\omega^{2}_{ci}-\omega^{2}\left(\mathbf{k}\right)\right)}\right].
\label{55}
\end{eqnarray}
\end{widetext}

It follows from Eqs. (\ref{52}) and (\ref{53}) that the average ion flow with velocities $\bar{U}_{ix}\left(X_{i}\right)$ and 
$\bar{U}_{iy}\left(X_{i}\right)$ stem from the spatial non-uniformity of the IC parametric turbulence. It develops in the inhomogeneous plasmas under the 
action of the inhomogeneous FW in front of the antenna. The Vlasov equation for the electron distribution function $F_{e}\left(\mathbf{v}_{e}, 
X_{e}, Y_{e}, t\right)$, which determines the effect of the inhomogeneous IC turbulence on the electrons is given 
by Eqs. (\ref{49}) - (\ref{55}) with species index $i$ changed on $e$. It should be noted also, that the spectral intensity of the electrostatic potential is 
the same in the ion and the electron frames, i. e. $\left|\varphi^{(e)}_{e}\left(\mathbf{k}, 
X_{e}\right)\right|^{2}=\left|\varphi\left(\mathbf{k}, X_{i}\right)\right|^{2}$, as it follows from Eq. (\ref{21}). At the present time, this function may 
be determined experimentally or by numerical simulations for the given experimental conditions.  The simple estimates for the velocities $\bar{U}_{ix}
\left(X_{i}\right)$ and $\bar{U}_{iy}\left(X_{i}\right)$ of the ion flow driven by the inhomogeneous short scale electrostatic turbulence follow  from Eqs. 
(\ref{52}) and (\ref{53}) 
\begin{eqnarray}
&\displaystyle
\bar{U}_{ix}\left(X_{i}\right)\sim \bar{U}_{iy}\sim v_{Ti}\left(\frac{e\widetilde{\varphi}\left(X_{i}\right)}{T_{i}}\right)^{2}k^{2}_{0}\rho^{2}_{i}
\frac{\rho_{i}}{L},
\label{56}
\end{eqnarray}
where $L$ is the spatial inhomogeneity scale length of the spectral intensity.  The estimate similar to (\ref{56}) we derive for the electron flow 
velocities, 
\begin{eqnarray}
&\displaystyle
\bar{U}_{ex}\left(X_{i}\right)\sim \bar{U}_{ey}\left(X_{i}\right)\sim\bar{U}_{ix}\left(X_{i}\right).
\label{57}
\end{eqnarray}
For the estimate (\ref{32}) for the level of the IC turbulence Eq. (\ref{55}) gives
\begin{eqnarray}
&\displaystyle
\bar{U}_{ix}\left(x_{i}\right)\sim v_{Ti}\frac{\rho_{i}}{L}\left(\frac{U}{v_{Ti}}\right)^{3}.
\label{58}
\end{eqnarray}
where $v_{Ti}$ and $U$ are the functions of $X_{i}$. For the numerical sample, grounded on the numerical data from
Ref.\cite{Bertelli} and presented after Eq. (\ref{32}), $U\left(X_{i}\right) \approx v_{Ti}$, $\rho_{i}\approx 0.2$ cm and $L\lesssim 2$ cm 
(Fig. 1 in Ref. \cite{Bertelli}), $\bar{U}_{ix}\left(X_{i}\right)\sim v_{Ti}\left(\rho_{i}/L\right)\sim 0.1v_{Ti}$.

A plasma in front of the antenna has strong gradients of the density and of the ion temperature. In such a plasma, the low frequency, $\omega
\left(\mathbf{k}\right)\ll \omega_{ci}$, long wavelength, $k_{\bot}\rho_{i}\lesssim 1$, drift type instabilities develop which are at the origin 
of the drift turbulence. The spatial inhomogeneity of this turbulence, determined by the radial coordinate $X_{i}$, is also the source 
of the development of the plasma flows across the magnetic field.

On the time scale $t$ of the order of the period of the drift wave, $\omega^{-1}\left(\mathbf{k}\right)\gg \omega^{-1}_{ci}$, the averaged over time $t \gg 
\omega_{0}^{-1}$ velocities $U_{ix}\left(X_{i}, t\right)$ and $U_{iy}\left(X_{i}, t\right)$, determined by Eqs. (\ref{A5}), (\ref{A6}),  are equal to 
zero. The transition to the frame of references, which moves with a random velocity $\mathbf{\tilde{V}}_{i}\left(\mathbf{r}_{i}, t\right)$ of the ion motion 
in the fields of the drift turbulence, transforms the Vlasov equation (\ref{13}) to Eq. (\ref{49}) in which variable $X_{i}$ is determined in the 
laboratory frame. In this case of the drift turbulence the velocities $\bar{U}_{ix}\left(x_{i}, t\right)$ and $\bar{U}_{iy}\left(x_{i}, t\right)$ 
are determined by Eqs. (\ref{52}), (\ref{53}), in which functions $a_{i1}\left(\mathbf{k}\right)$ and $a_{i1}\left(\mathbf{k}\right)$ for $\omega
\left(\mathbf{k}\right)\ll \omega_{ci}$ are equal approximately to
\begin{eqnarray}
&\displaystyle
a_{i1}\left(\mathbf{k}\right)\approx \frac{1}{\omega\left(\mathbf{k}\right)\omega_{ci}},
\qquad a_{i2}\left(\mathbf{k}\right)\approx \frac{1}{\omega^{2}_{ci}}\ll a_{i1}.
\label{59}
\end{eqnarray} 
Therefore, the estimates for the velocities $\bar{U}_{ix}\left(x_{i}\right), \bar{U}_{iy}\left(x_{i}\right)$ of the flow driven by the drift turbulence 
inhomogeneity are
\begin{eqnarray}
&\displaystyle
\bar{U}_{ix}\left(X_{i}\right)\sim \bar{U}_{iy}\left(X_{i}\right)\sim v_{Ti}k_{0\bot}^{2}\rho^{2}_{i}\left(\frac{e\widetilde{\varphi}}{T_{i}}\right)^{2}
\frac{v_{Ti}}{\omega\left(\mathbf{k}_{0}\right)L}.
\label{60}
\end{eqnarray} 
Similar estimates were derived for the velocities of the electron flow, driven by the spatial inhomogeneity of the drift turbulence,
\begin{eqnarray}
&\displaystyle
\bar{U}_{ex}\left(X_{i}\right)\sim \bar{U}_{ey}\left(X_{i}\right)\sim \bar{U}_{ix}\left(X_{i}\right).
\label{61}
\end{eqnarray} 

Equation (\ref{49}) is the basic equation for the derivation, as the corresponding moments of that equation,
the equations which determines the mesoscale evolution of the plasma in the region of the strong inhomogeneity of the plasma and FW. By
multiplying Eq. (\ref{49}) on $m_{i}v^{2}_{i\bot}/2$ and integrating this equation over velocities $\mathbf{v}_{i}$, we obtain the
equation 
\begin{eqnarray}
&\displaystyle \frac{\partial n_{0i}T_{i\bot}}{\partial t}-\bar{U}_{ix}\left(X_{i}\right)\frac{\partial n_{0i}T_{i\bot}}{\partial X_{i}}
\nonumber\\ 
& \displaystyle
-\bar{U}_{iy}\left(X_{i}\right)\frac{\partial n_{0i}T_{i\bot}}{\partial Y_{i}}=0,
\label{62}
\end{eqnarray} 
which governs the temporal evolution of the averaged density of the perpendicular thermal energy of ions, $n_{0i}\left(X_{i}, Y_{i}, t\right)
T_{i\bot}\left(X_{i}, Y_{i}, t\right)$, resulted from the interaction of ions with spatially non-uniform short scale IC and drift turbulence.  
The terms  with velocities $\bar{U}_{ix}\left(x_{i}\right)$ 
and $\bar{U}_{iy}\left(X_{i}\right)$ in Eq. (\ref{62}) determine the convective transport of the ion thermal energy in the direction 
of $-X_{i}$ and $-Y_{i}$, originated from the IC and drift turbulence inhomogeneity along $X_{i}$.

The simple solution to Eq. (\ref{62}) can be derived for the case of the spatially uniform velocities $\bar{U}_{ix}$ and $\bar{U}_{iy}$. 
We derive the solution for this case as of the initial value problem with initial condition that at $t=t_{0}$
\begin{eqnarray}
&\displaystyle n_{0i}\left(X_{i}, Y_{i}, t_{0}\right)T_{i\bot}\left(X_{i}, Y_{i}, t_{0}\right)= \Psi_{i}\left(X_{i}, Y_{i}\right).
\label{63}
\end{eqnarray}
The equations of characteristics for Eq. (\ref{61}),
\begin{eqnarray}
&\displaystyle dt=\frac{dX_{i}}{-\bar{U}_{ix}}=\frac{dY_{i}}{-\bar{U}_{iy}},
\label{64}
\end{eqnarray}
have two integrals 
\begin{eqnarray}
&\displaystyle X_{i0}=X_{i}+\bar{U}_{ix}\left(t-t_{0}\right),
\nonumber\\ 
&\displaystyle
Y_{i0}=Y_{i}+\bar{U}_{iy}\left(t-t_{0}\right),
\label{65}
\end{eqnarray}
which correspond to the initial conditions $X_{i}=X_{i0}$ and $Y_{i}=Y_{i0}$ at $t=t_{0}$. Then, the solution to the initial value problem 
(\ref{62}), (\ref{63})  are 
\begin{eqnarray}
&\displaystyle 
n_{0i}\left(X_{i}, Y_{i}, t\right)T_{i\bot}\left(X_{i}, Y_{i}, t\right)
\nonumber\\ 
&\displaystyle
= \Psi_{0i}\left(X_{i}+\bar{U}_{ix}\left(t-t_{0}\right), Y_{i}+\bar{U}_{iy}\left(t-t_{0}\right)\right)
\label{66}
\end{eqnarray}
Equation (\ref{66}) demonstrates that the ion thermal energy density 
$n_{0i}\left(X_{i}, Y_{i}, t\right)T_{i\bot}\left(X_{i}, Y_{i}, t\right)$, which was equal to $\Psi_{i}\left(X_{i0}, Y_{i0}\right)$ in point 
$X_{i}=X_{i0}$ and $Y_{i}=Y_{i0}$ in the pedestal at time $t_{0}$,  transported outward of LCMS to the SOL region with radial velocity $\bar{U}_{ix}$ and 
poloidal velocity $\bar{U}_{iy}$ in position $X_{i}$, $Y_{i}$ at time $t$.  This thermal energy flux is limited along $X_{i}$ by the region of SOL, where 
the strong gradients of the plasma parameters and of FW electric field exist; it has low radial and poloidal velocities in the far SOL region with much more 
uniform plasma parameters. 

It is instructive to present the estimate for the ion thermal energy flux 
\begin{eqnarray}
&\displaystyle
J_{ix}\sim n_{i0}T_{i}\bar{U}_{ix},
\label{67}
\end{eqnarray}
delivered from the pedestal to SOL by the turbulent convection. For the plasma density $n_{0i}=10^{19}$\,m$^{-3}$, the ion temperature $T_{i}= 100$ eV 
in the pedestal plasma and $\bar{U}_{ix}=0.1v_{Ti}$, estimate (\ref{67}) gives $J_{ix}\sim 1,5$ MWm$^{-2}$ ion heat flux to SOL from the pedestal 
region. The estimate for the electron thermal energy flux with $T_{e}\approx T_{i}$ and $\bar{U}_{ix}\approx \bar{U}_{ex}$ is on the same order. 

This result indicates that a significant part of the FW power deposited to the edge layers of the inhomogeneous plasma inside LCMS is 
convected to SOL by the flow caused by the inhomogeneity of the short-scale IC and drift turbulence. This flow continues in SOL as a flow, which 
transport of a plasma almost along field line with small poloidal deviation from the field line caused by the poloidal velocity $\bar{U}_{iy}$.

\section{Conclusions}\label{sec5}

In this paper, we present the two-scale approach to the theory of the IC parametric turbulence, driven by the strong inhomogeneous FW electric field, 
which develops in the inhomogeneous plasmas. This approach reveals the effect of the formation of the mesoscale convective 
flow of such a plasma caused by the spatial inhomogeneity of the microscale IC or drift turbulence. The radial and poloidal velocities of this
flow are proportional to the gradient of the spectral intensity of the small-scale turbulence, directed along the radial coordinate $x_{i}$.  
The radial flow velocity, $\bar{U}_{ix}$, is in the direction opposite to this gradient. 

In the tokamak plasmas, this flow moves outward of pedestal to the SOL region with radial velocity 
$\bar{U}_{ix}\left(X_{i}\right)$ and with poloidal velocity $\bar{U}_{iy}\left(X_{i}\right)$. In SOL, this flow is bounded along $x_{i}$ by the region
where the strong gradients of the plasma parameters and of FW electric field exist and flows almost along magnetic field lines 
to the divertor plates. We found that this convective flow may be responsible for the transport of the plasma thermal energy from 
the strongly inhomogeneous high density hot pedestal plasma to the lower density cold  near-SOL plasma. This result explains the origin in SOL 
of the significant fraction of the FW power applied to NSTX, which is delivered eventually along magnetic field line by SOL flow to the divertor
\cite{Perkins,Perkins1}. 

It should be note, that the derived kinetic equation (\ref{49}) with the mesoscale flow velocities $\bar{U}_{ix}\left(X_{i}\right)$ 
and $\bar{U}_{iy}\left(X_{i}\right)$, determined by Eqs. (\ref{50}), (\ref{51}), which governs the mesoscale evolution of the ion distribution function, 
has a rather general form for spatially inhomogeneous plasma microturbulence. Therefore, it may be concluded 
that the effect of the mesoscale convective flows formation is likely the inherent effect caused by the spatially inhomogeneous microscale 
plasma turbulence.

\begin{acknowledgments}
This work was supported by National R\&D Program through the National Research Foundation of 
Korea (NRF) funded by the Ministry of Education, Science and Technology (Grant No. NRF-2018R1D1A3B07051247) and BK21 FOUR, 
the Creative Human Resource Education and Research Programs for ICT Convergence in the 4th Industrial Revolution.
\end{acknowledgments}

\bigskip
{\bf DATA AVAILABILITY}

\bigskip
The data that support the findings of this study are available from the corresponding author upon reasonable request.

\appendix 
\section{{Solutions to Eqs. (\ref{10}) and (\ref{11}) for $V_{ix}\left(X_{i}, t\right)$ and $V_{iy}\left(X_{i}, t\right)$}}
In this Appendix, we present the approximate solution to Eqs. (\ref{10}), (\ref{11}) for  $V_{ix}\left(X_{i}, t\right)$ and 
$V_{iy}\left(X_{i}, t\right)$ in the limit of small displacement $|R_{i}\left(X_{i}, t\right)|$ of ion in the inhomogeneous FW field compared 
with the electric field inhomogeneity scale length $L_{E}$. Here, we employ the procedure developed in Ref.\cite{Mikhailenko3} for approximate
solution of equations, similar to Eqs. (\ref{10}) and (\ref{11}), when FW field has a structure of a plane wave  with a finite wave length. 
Direct integration of Eqs. (\ref{10}), (\ref{11}) gives
\begin{widetext}
\begin{eqnarray}
&\displaystyle
V_{ix}\left(X_{i}, t\right)=\frac{e}{2m_{i}}\cos \omega_{ci}t\int\limits ^{t}_{0}dt_{1}E_{0x}\left(X_{i}+R_{i}\left(X_{i}, t\right)\right)
\left[\cos \left(\left(\omega_{0}+\omega_{ci}\right)t_{1}\right)+\cos\left(\left(\omega_{0}-\omega_{ci}\right)t_{1}\right) \right] 
\nonumber\\ 
&\displaystyle
-\frac{e}{2m_{i}}\cos \omega_{ci}t\int\limits ^{t}_{0}dt_{1}E_{0y}\left(X_{i}+R_{i}\left(X_{i}, t\right)\right)
\left[\cos \left(\left(\omega_{0}-\omega_{ci}\right)t_{1}\right)-\cos\left(\left(\omega_{0}+\omega_{ci}\right)t_{1}\right) \right] 
\nonumber\\ 
&\displaystyle
+\frac{e}{2m_{i}}\sin \omega_{ci}t\int\limits ^{t}_{0}dt_{1}E_{0x}\left(X_{i}+R_{i}\left(X_{i}, t\right)\right)
\left[\sin \left(\left(\omega_{0}+\omega_{ci}\right)t_{1}\right)-\sin\left(\left(\omega_{0}-\omega_{ci}\right)t_{1}\right) \right] 
\nonumber\\ 
&\displaystyle
+\frac{e}{2m_{i}}\sin\omega_{ci}t\int\limits ^{t}_{0}dt_{1}E_{0y}\left(X_{i}+R_{i}\left(X_{i}, t\right)\right)
\left[\sin\left(\left(\omega_{0}+\omega_{ci}\right)t_{1}\right)+\sin\left(\left(\omega_{0}-\omega_{ci}\right)t_{1}\right) \right]
\nonumber\\ 
&\displaystyle
-\frac{e}{m_{i}\omega_{0}}E'_{0y}\left(X_{i}\right)\int\limits ^{t}_{0}dt_{1}\cos \omega_{0}t_{1}\left[V_{ix}\left(X_{i}, t_{1}\right)\sin \omega_{ci}
\left(t-t_{1}\right)- V_{iy}\left(X_{i}, t_{1}\right)\cos \omega_{ci}\left(t-t_{1}\right)\right]  
\label{A1}
\end{eqnarray}
and 
\begin{eqnarray}
&\displaystyle
V_{iy}\left(X_{i}, t\right)=-\frac{e}{2m_{i}}\sin \omega_{ci}t\int\limits ^{t}_{0}dt_{1}E_{0x}\left(X_{i}+R_{i}\left(X_{i}, t\right)\right)
\left[\cos \left(\left(\omega_{0}+\omega_{ci}\right)t_{1}\right)+\cos\left(\left(\omega_{0}-\omega_{ci}\right)t_{1}\right) \right] 
\nonumber\\ 
&\displaystyle
+\frac{e}{2m_{i}}\sin \omega_{ci}t\int\limits ^{t}_{0}dt_{1}E_{0y}\left(X_{i}+R_{i}\left(X_{i}, t\right)\right)
\left[\cos \left(\left(\omega_{0}-\omega_{ci}\right)t_{1}\right)-\cos\left(\left(\omega_{0}+\omega_{ci}\right)t_{1}\right) \right] 
\nonumber\\ 
&\displaystyle
+\frac{e}{2m_{i}}\cos \omega_{ci}t\int\limits ^{t}_{0}dt_{1}E_{0x}\left(X_{i}+R_{i}\left(X_{i}, t\right)\right)
\left[\sin \left(\left(\omega_{0}+\omega_{ci}\right)t_{1}\right)-\sin\left(\left(\omega_{0}-\omega_{ci}\right)t_{1}\right) \right] 
\nonumber\\ 
&\displaystyle
+\frac{e}{2m_{i}}\cos\omega_{ci}t\int\limits ^{t}_{0}dt_{1}E_{0y}\left(X_{i}+R_{i}\left(X_{i}, t\right)\right)
\left[\sin\left(\left(\omega_{0}+\omega_{ci}\right)t_{1}\right)+\sin\left(\left(\omega_{0}-\omega_{ci}\right)t_{1}\right) \right] 
\nonumber\\ 
&\displaystyle
-\frac{e}{m_{i}\omega_{0}}E'_{0y}\left(X_{i}\right)\int\limits ^{t}_{0}dt_{1}\cos \omega_{0}t_{1}\left[V_{ix}\left(X_{i}, t_{1}\right)\cos \omega_{ci}
\left(t-t_{1}\right)+V_{iy}\left(X_{i}, t_{1}\right)\sin\omega_{ci}\left(t-t_{1}\right)\right]  
\label{A2}
\end{eqnarray}
\end{widetext}
By partial integration of Eqs. (\ref{A1}) and (\ref{A2}), a power series expansion in powers of 
$\left| R_{i}\left(X_{i}, t\right), t\right)/L_{E}|<1$ can be derived for $V_{ix}\left(X_{i}, t\right)$ and $V_{iy}\left(X_{i}, t\right)$, assuming that $
\omega_{0}\sim \omega_{ci}\sim |\omega_{0}-\omega_{ci}|$. 
We obtain on this way, neglecting by the terms on the order of $O\left(\left| R_{i}\left(X_{i}, t\right)|/L_{E}\right)^{2}\right)$ and above, that
\begin{eqnarray}
&\displaystyle
V_{ix}\left(X_{i}, t\right)=U_{ix}\left(X_{i}, t\right)
\nonumber\\ 
&\displaystyle
\times\left[1+\frac{e_{i}}{m_{i}}\frac{\cos \omega_{0}t}{\left(\omega^{2}_{0}-\omega^{2}_{ci}\right)^{2}}\left(E'_{0x}\left(X_{i}\right)
\left(\omega^{2}_{0}+\omega^{2}_{ci}\right)\right.\right.
\nonumber\\ 
&\displaystyle
\left.\left.-2E'_{0y}\left(X_{i}\right)\omega_{0}\omega_{ci}\right)\right]
+\frac{e}{2m_{i}\omega_{0}}\frac{E'_{0y}\left(X_{i}\right)\sin 2\omega_{0}t}{\left(4\omega^{2}_{0}-\omega^{2}_{ci}\right)}
\nonumber\\ 
&\displaystyle
\times\left(2\omega_{0}U_{iy}\left(X_{i}\right)+\omega_{ci}U_{ix}\left(X_{i}\right)\right),
\label{A3}
\end{eqnarray}
and
\begin{eqnarray}
&\displaystyle
V_{iy}\left(X_{i}, t\right)=U_{iy}\left(X_{i}, t\right)
\nonumber\\ 
&\displaystyle
+\frac{e_{i}}{m_{i}}U_{ix}\left(X_{i}, t\right)\frac{\sin \omega_{0}t}{\left(\omega^{2}_{0}-\omega^{2}_{ci}\right)^{2}}\left(-2E'_{0x}\left(X_{i}\right)
\omega_{0}\omega_{ci}\right.
\nonumber\\ 
&\displaystyle
\left.+E'_{0y}\left(X_{i}\right)\left(\omega^{2}_{0}+\omega^{2}_{ci}\right)\right)
-\frac{eE'_{0y}\left(X_{i}\right)}{2m_{i}\omega_{0}\omega_{ci}}U_{iy}\left(X_{i}\right)
\nonumber\\ 
&\displaystyle
-\frac{eE'_{0y}\left(X_{i}\right)}{2m_{i}\omega_{0}}
\frac{\cos 2\omega_{0}t}{\left(4\omega_{0}^{2}-\omega^{2}_{ci}\right)}\left(\omega_{ci}U_{iy}\left(X_{i}\right)+2\omega_{0}U_{ix}\left(X_{i}\right)\right),
\label{A4}
\end{eqnarray}
where the prime in $E'_{0x}\left(X_{i}\right)$ denotes the derivative with respect to the spatial coordinate $X_{i}$ and
\begin{eqnarray}
&\displaystyle
U_{ix}\left(X_{i}, t\right)=\frac{e_{i}}{m_{i}\left(\omega^{2}_{0}-\omega^{2}_{ci}\right)}
\nonumber\\ 
&\displaystyle
\times\left(\omega_{0}E_{0x}\left(X_{i}\right)-\omega_{ci}E_{0y}\left(X_{i}\right)\right)\sin \omega_{0}t
\nonumber 
\\
&\displaystyle
=U_{ix}\left(X_{i}\right)\sin\omega_{0}t,
\label{A5}
\end{eqnarray}
and
\begin{eqnarray}
&\displaystyle
U_{iy}\left(X_{i}, t\right)=\frac{e_{i}}{m_{i}\left(\omega^{2}_{0}-\omega^{2}_{ci}\right)}
\nonumber\\ 
&\displaystyle
\times\left(\omega_{ci}E_{0x}\left(X_{i}\right)-\omega_{0}E_{0y}\left(X_{i}\right)\right)\cos \omega_{0}t
\nonumber 
\\
&\displaystyle
=U_{iy}\left(X_{i}\right)\cos\omega_{0}t.
\label{A6}
\end{eqnarray}
In should be noted that the developed expansion procedure is divergent and gives wrong result for the condition of the IC resonance for which $\omega_{0}
\approx\omega_{ci}\gg |\omega_{0}-\omega_{ci}|$. For $\omega_{0}\sim \omega_{ci}\sim |\omega_{0}-\omega_{ci}|$ the estimate
\begin{eqnarray}
&\displaystyle 
\frac{eE'_{0x,0y}\left(X_{i}\right)}{m_{i}}\frac{\left(\omega^{2}_{0}+\omega^{2}_{ci}\right)}{\left(\omega^{2}_{0}-\omega^{2}_{ci}\right)^{2}}\sim 
\frac{cE_{0x,0y}}{\omega_{ci}B_{0}}\frac{1}{L_{E}}\sim \frac{\xi}{L_{E}}
\label{A7}
\end{eqnarray}
follows, where $\xi$ is the ion displacement in the $E_{0x}$ FW field over time $\sim \omega^{-1}_{0}\sim \omega^{-1}_{ci}$. The same estimates follow for 
other $E'_{0x}\left(X_{i}\right)$ and $E'_{0y}\left(X_{i}\right)$ contained terms in Eqs. (\ref{A1}) and (\ref{A2}). For the numerical data of Ref.
\cite{Bertelli, Bertelli1} ($\omega_{0}=1,8\cdot 10^{8}$s$^{-1}$ $\left(f_{0}=30 \,{\text {MHz}}\right)$, $\omega_{ci} \approx 10^{8}$ s$^{-1}$ 
$\left(B_{0}=1\,{\text{T}}\right)$,
$E_{0x}=450$ V/cm and $L_{E}=1$cm), we have $\xi/L_{E}\sim 2.5\cdot 10^{-2}\ll 1$. Thus, the effect of the spatial inhomogeneity of $E_{0x}$ on the 
velocities $V_{ix}\left(X_{i}, t\right)$ and $V_{iy}\left(X_{i}, t\right)$ is small and the approximations 
\begin{eqnarray}
&\displaystyle
V_{ix}\left(X_{i}, t\right)\approx U_{ix}\left(X_{i}, t\right), \,\,\,\,V_{iy}\left(X_{i}, t\right)\approx U_{iy}\left(X_{i}, t\right)
\label{A8}
\end{eqnarray}
will be used in what follows. 

With velocities $V_{ix}\left(X_{i}, t\right)$ and $V_{iy}\left(X_{i}, t\right)$, determined by Eq. (\ref{A8}), the Vlasov equation (\ref{6}) 
for the ion distribution function $F_{i}\left(\mathbf{v}_{i}, \mathbf{r}_{i}, t\right)$ becomes
\begin{widetext}
\begin{eqnarray}
&\displaystyle
\frac{\partial F_{i}\left(\mathbf{v}_{i}, \mathbf{r}_{i}, X_{i}, t\right)}{\partial t}+\left[v_{ix}\left(1+\frac{e}{m_{i}}\frac{\cos \omega_{0}t}{\omega_{0}
\left(\omega^{2}_{0}-\omega^{2}_{ci}\right)}\left(\omega_{0}E'_{0x}\left(X_{i}\right)-\omega_{ci}E'_{0y}\left(X_{i}\right)\right)\right)\right.
\nonumber\\ 
&\displaystyle
\left.+\frac{e^{2}}{4m^{2}_{i}}\frac{\sin 2\omega_{0}t}{\omega_{0}\left(\omega^{2}_{0}-\omega^{2}_{ci}\right)^{2}}
\left(\omega^{2}_{0}\frac{dE^{2}_{0x}\left(X_{i}\right)}{dX_{i}}+\omega^{2}_{ci}\frac{dE^{2}_{0y}\left(X_{i}\right)}{dX_{i}}
-\omega_{0}\omega_{ci}\frac{d\left(E_{0x}\left(X_{i}\right)E_{0y}\left(X_{i}\right)\right)}{dX_{i}}\right)\right]
\frac{\partial F_{i}}{\partial x_{i}} 
\nonumber\\ 
&\displaystyle
+\left[v_{iy}-v_{ix}\frac{e}{m_{i}}\frac{\sin \omega_{0}t}{\omega_{0}
\left(\omega^{2}_{0}-\omega^{2}_{ci}\right)}\left(\omega_{ci}E'_{0x}\left(X_{i}\right)-\omega_{0}E'_{0y}\left(X_{i}\right)\right)\right.
\nonumber\\ 
&\displaystyle
\left.-\frac{e^{2}}{4m^{2}_{i}}\frac{\left(1-\cos 2\omega_{0}t\right)}{\omega_{0}\left(\omega^{2}_{0}-\omega^{2}_{ci}\right)^{2}}
\left(\omega_{0}\omega_{ci}\frac{d}{dX_{i}}\left(E^{2}_{0x}\left(X_{i}\right)+E^{2}_{0y}\left(X_{i}\right)\right)-\left(\omega_{0}^{2}
+\omega^{2}_{ci}\right)\frac{d\left(E_{0x}\left(X_{i}\right)E_{0y}\left(X_{i}\right)\right)}{dX_{i}}\right)\right]
\frac{\partial F_{i}}{\partial y_{i}}+v_{iz}\frac{\partial F_{i}}{\partial z_{i}} 
\nonumber\\ 
&\displaystyle
+\omega_{ci}\left(v_{iy}\frac{\partial F_{i}}{\partial v_{ix}}-v_{ix}\frac{\partial F_{i}}{\partial v_{iy}}\right)+\left[v_{iy}\frac{e}{m_{i}\omega_{0}}E'_{0y}\left(X_{i}\right)\cos \omega_{0}t
-v_{ix}\frac{e}{m_{i}}\frac{\sin\omega_{0}t}{\left(\omega^{2}_{0}-\omega^{2}_{ci}\right)}\left(\omega_{0}E'_{0x}\left(X_{i}\right)-\omega_{ci}
E'_{0y}\left(X_{i}\right)\right)\right] \frac{\partial F_{i}}{\partial v_{ix}}
\nonumber\\ 
&\displaystyle
-v_{ix}\frac{e}{m_{i}}\cos \omega_{0}t\left[\frac{1}{\omega_{0}}E'_{0y}\left(X_{i}\right)+\frac{1}{\left(\omega^{2}_{0}-\omega^{2}_{ci}\right)}
\left(\omega_{ci}E'_{0x}\left(X_{i}\right)-\omega_{0}E'_{0y}\left(X_{i}\right)\right)\right]\frac{\partial F_{i}}{\partial v_{iy}}
-\frac{e_{i}}{m_{i}}\nabla\varphi\left(\mathbf{r}_{i}, t\right)
\frac{\partial F_{i}\left(\mathbf{v}_{i}, \mathbf{r}_{i}, X_{i}, t\right)}{\partial \mathbf{v}_{i}}=0.
\label{A9}
\end{eqnarray}
\end{widetext}
It follows from Eq. (\ref{A9}), that the Vlasov equation in the ion oscillating frame contains only the derivatives $E'_{0x}\left(X_{i}\right)$ 
and $E'_{0y}\left(X_{i}\right)$ of FW electric field components. All these terms are of the order of $\xi/L_{E}\sim 2.5\cdot 10^{-2}\ll 1$ for the numerical 
data considered above and should be neglected. Without these terms, Eq. (\ref{A9}) has a form (\ref{13}) as for the plasma without the FW field.

{}

\begin{thebibliography}{}

\bibitem{Stix} T.~H.~Stix, "Fast-wave heating of a two-component plasma",  Nucl. Fusion {\bf 15}, 737 (1975).

\bibitem{Hosea} J.~Hosea, R.~E.~Bell, B.~P.~LeBlanc, C.~K.~Phillips, G.~Taylor, E.~Valeo, J.~R.~Wilson, E.~F.~Jaeger, P.~M.~Ryan, 
J.~Wilgen, H.~Yuh, F.~Levinton, S.~Sabbagh, K.~Tritz, J.~Parker, P.~T.~Bonoli, R.~Harvey, 
and NSTX Team, "High harmonic fast wave heating efficiency enhancement and current drive
at longer wavelength on the National Spherical Torus Experiment," Phys. Plasmas {\bf 15}, 056104 (2008).

\bibitem{Prater} R.~Prater, C.~P.~Moeller, R.~I.~Pinsker, M.~Porkolab, O.~Meneghini, V.~L.~Vdovin, "Application of very high 
harmonic fast waves for off-axis current drive in the DIII-D and FNSF-AT tokamaks," Nucl. Fusion {\bf 54}, 083024 (2014). 

\bibitem{Pinsker} R.~I.~Pinsker, R.~Prater, C.~P.~Moeller, J.~S.~de Grassie, C.~C.~Petty, M.~Porkolab, J.~P.~Anderson, A.~M.~Garofalo, C.~Lau, A.~Nagy, 
D.~C.~Pace, H. Torreblanca, J.~G.~Watkins, L.~Zeng, "Experiments on helicons in DIII-D—investigation of
the physics of a reactor-relevant non-inductive current drive technology," Nucl. Fusion {\bf 58}, 106007 (2018). 

\bibitem{Wang} S.~J.~Wang, H.~H.~Wi, H.~J.~Kim, J.~Kim, J.~H.~Jeong and J.~G.~Kwak, "Helicon wave coupling in KSTAR plasmas for offaxis
current drive in high electron pressure plasmas," Nuclear Fusion {\bf 57}, 046010 (2017). 

\bibitem{Rost} J.~C.~Rost, M.~Porkolab, R.~L.~Boivin, "Edge ion heating and parametric decay during injection of ion cyclotron
resonance frequency power on the Alcator C-Mod tokamak," Phys. Plasmas {\bf 9}, 1262 (2002). 

\bibitem{Pace}D.~C.~Pace, R.~I.~Pinsker, W.~W.~Heidbrink, R.~K.~Fisher, M.~A.~Van~Zeeland, M.~E.~Austin, G.~R.~McKee, and
M. Garc$\acute{\text{i}}$a-Mu$\tilde{\text{n}}$oz, "Scrape-off layer ion acceleration during fast wave injection in the DIII-D tokamak", 
Nucl. Fusion {\bf 52}, 063019 (2012).

\bibitem{Wilson} J.~R.~Wilson, S.~Bernabei, T.~Biewer, S.~Diem, J.~Hosea, B.~LeBlanc, C.~K.~ 
Phillips, P.~Ryan, and D.~W.~Swain. "Parametric Decay During HHFW on NSTX". AIP Conference Proceedings {\bf 787}, 66 (2005).

\bibitem{Porkolab1} M.~Porkolab, "Parametric processes in magnetically confined CTR plasmas," Nuclear Fusion {\bf 18}, 367 (1978).

\bibitem{Porkolab2} M.~Porkolab, "Parametric instabilities in the tokamak edge plasma in the ion cyclotron heating regimes," 
Fusion Engineering and Design {\bf 12}, 93 (1990).

\bibitem{Mikhailenko1} V.~S.~Mikhailenko, K.~N.~Stepanov, "Theory of weak parametric plasma turbulence," 
Zh.~Eksp.~Teor.~Fiz. {\bf 87}, 161 (1984)[Sov. Phys. JETP {\bf 60},  92 (1984)].

\bibitem{Mikhailenko2} V.~S.~Mikhailenko, E.~E.~Scime, "Effect of ion cyclotron parametric turbulence on the generation of edge 
suprathermal ions during ion cyclotron plasma heating," Phys. Plasmas {\bf 11}, 3691 (2004).

\bibitem{Mikhailenko3} V.~S.~Mikhailenko, V.~V.~Mikhailenko, Hae~June~Lee, "The ion cyclotron parametric instabilities and 
the anomalous heating of ions in the tokamak edge plasma in the fast wave heating regime," Phys. Plasmas {\bf 27}, 052508  (2020).

\bibitem{Perkins} R.~J.~Perkins, J.~C.~Hosea, G.~J.~Kramer, J.~-W.~Ahn, R.~E.~Bell, A.~Diallo, S.~Gerhardt,
T.~K.~Gray, D.~L.~Green, E.~F.~Jaeger, M.~A.~Jaworski, B.~P.~LeBlanc, A.~McLean, R.~Maingi, C.~K.~Phillips, L.~Roquemore, 
P.~M.~Ryan, S.~Sabbagh, G. Taylor, J.~R.~Wilson, "High-Harmonic Fast-Wave Power Flow along Magnetic Field Lines in the Scrape-Off 
Layer of NSTX," Phys. Rev. Lett. {\bf 109}, 045001 (2012).

\bibitem{Perkins1} R.~J.~Perkins, J.~-W.~Ahn, R.~E.~Bell, A.~Diallo, S.~Gerhardt, T.~K.~Gray, D.~L.~Green, E.~F.~Jaeger, J.~C.~Hosea, 
M.~A.~Jaworski, B.~P.~LeBlanc, G.~J.~Kramer, A.~McLean, R.~Maingi, C.~K.~Phillips, M.~Podest`a, L.~Roquemore, 
P.~M.~Ryan, S.~Sabbagh, F.~Scotti, G.~Taylor, J.~R.~Wilson, "Fast-wave power flow along SOL field lines in NSTX and the associated 
power deposition profile across the SOL in front of the antenna," Nucl. Fusion {\bf 53}, 083025 (2013). 

\bibitem{Jaeger} E. F. Jaeger, L. A. Berry, E. D’Azevedo, D. B. Batchelor, and M. D. Carter, "All-orders spectral calculation of 
radio-frequency heating in two-dimensional toroidal plasmas", Phys. Plasmas {\bf 8}, 1573 (2001).

\bibitem{Green} D.~L.~Green, L.~A.~Berry, G.~Chen, P.~M.~Ryan, J.~M.~Canik,  E.~F.~Jaeger, "Predicting High Harmonic Ion Cyclotron 
Heating Efficiency in Tokamak Plasmas", Phys. Rev. Lett. {\bf 107}, 145001 (2011).

\bibitem{Bertelli} N.~Bertelli, E.~F.~Jaeger, J.~C.~Hosea, C.~K.~Phillips, L.~Berry, S.~P.~Gerhardt, D.~Green, 
B.~LeBlanc, R.~J.~Perkins, P.~M.~Ryan, G.~Taylor, E.~J.~Valeo, J.~R.~Wilson, "Full wave simulations of fast wave heating losses
in the scrape-off layer of NSTX and NSTX-U," Nucl. Fusion {\bf 54}, 083004 (2014). 

\bibitem{Bertelli1} N.~Bertelli, E.~F.~Jaeger, J.~C.~Hosea, C.~K.~Phillips, L.~Berry, P.~T.~Bonoli, S.~P.~Gerhardt, D.~Green, 
B.~LeBlanc, R.~J.~Perkins, C.~M.~Qin, R.~I.~Pinsker, R.~Prater, P.~M.~Ryan, G.~Taylor, E.~J.~Valeo, J.~R.~Wilson, 
J~.C.~Wright, X.~J.~Zhang, "Full wave simulations of fast wave efficiency and power losses in the scrape-off layer of tokamak
plasmas in mid/high harmonic and minority heating regimes," Nucl. Fusion {\bf 56}, 016019 (2016). 


\bibitem{Vdovin} V.~L.~Vdovin, "Electromagnetic theory of an antenna for ICR heating of tokamak plasmas,"  Nucl. Fusion {\bf 23}, 1435 (2083).

\bibitem{Dum} C.~T.~Dum, T.~H.~Dupree, "Nonlinear stabilization of high-frequency instabilities in a magnetic field," 
Phys. Fluids {\bf 13}, 2064 (1971).

\end{thebibliography}
\end{document}